\documentclass[a4paper,12pt]{article}
\pdfoutput=1
\usepackage{epsfig}
\usepackage{amssymb}
\usepackage{amsfonts}
\usepackage{amsmath}
\usepackage{euscript}
\usepackage{verbatim}
\usepackage{latexsym}
\usepackage{graphicx}
\usepackage{caption}
\usepackage{breqn}
\usepackage{float}
\usepackage{subcaption}
\usepackage{placeins}

\usepackage{listings}

\newif\ifdtup

\catcode`\@=11

\@addtoreset{equation}{section}

\def\@normalsize{\@setsize\normalsize{15pt}\xiipt\@xiipt
\abovedisplayskip 14pt plus3pt minus3pt%
\belowdisplayskip \abovedisplayskip
\abovedisplayshortskip \z@ plus3pt%
\belowdisplayshortskip 7pt plus3.5pt minus0pt}

\def\small{\@setsize\small{13.6pt}\xipt\@xipt
\abovedisplayskip 13pt plus3pt minus3pt%
\belowdisplayskip \abovedisplayskip
\abovedisplayshortskip \z@ plus3pt%
\belowdisplayshortskip 7pt plus3.5pt minus0pt
\def\@listi{\parsep 4.5pt plus 2pt minus 1pt
     \itemsep \parsep
     \topsep 9pt plus 3pt minus 3pt}}

\relax

\catcode`@=12

\catcode`\@=11

\def\section{\@startsection{section}{1}{\z@}{3.5ex plus 1ex minus
   .2ex}{2.3ex plus .2ex}{\large\bf}}

\def\SymBoxes#1#2#3#4{\newdimen\un@t \un@t#3%
\raisebox{#1}{\rule{#2\un@t}{#4}\hskip-#2\un@t
\@tempdimb\un@t \advance\@tempdimb by-#4\@tempcntb#2\relax%
\@whilenum{\@tempcntb>0}\do{
\rule{#4}{\un@t}\hskip\@tempdimb \advance\@tempcntb by\m@ne}%
\hskip-#2\un@t \rule[\un@t]{#2\un@t}{#4}%
\rule[\un@t]{#4}{#4}\hskip-#4
\rule{#4}{\un@t}}\hskip-#4}                

\begin{document}

\newcommand{\beq}{\begin{equation}}
\newcommand{\eeq}{\end{equation}}
\newcommand{\bea}{\begin{eqnarray}}
\newcommand{\eea}{\end{eqnarray}}
\newcommand{\beas}{\begin{eqnarray*}}
\newcommand{\eeas}{\end{eqnarray*}}
\newcommand{\defi}{\stackrel{\rm def}{=}}
\newcommand{\non}{\nonumber}
\newcommand{\bquo}{\begin{quote}}
\newcommand{\enqu}{\end{quote}}
\newcommand*{\Resize}[2]{\resizebox{#1}{!}{$#2$}}%
\renewcommand{\(}{\begin{equation}}
\renewcommand{\)}{\end{equation}}
\def \eqn#1#2{\begin{equation}#2\label{#1}\end{equation}}

\def\e{\epsilon}
\def\IZ{{\mathbb Z}}
\def\IR{{\mathbb R}}
\def\IC{{\mathbb C}}
\def\IQ{{\mathbb Q}}
\def\de{\partial}
\def\Tr{ \hbox{\rm Tr}}
\def\H{ \hbox{\rm H}}
\def\HE{ \hbox{$\rm H^{even}$}}
\def\HO{ \hbox{$\rm H^{odd}$}}
\def\K{ \hbox{\rm K}}
\def\Im{ \hbox{\rm Im}}
\def\Ker{ \hbox{\rm Ker}}
\def\const{\hbox {\rm const.}}
\def\o{\over}
\def\im{\hbox{\rm Im}}
\def\re{\hbox{\rm Re}}
\def\bra{\langle}\def\ket{\rangle}
\def\Arg{\hbox {\rm Arg}}
\def\Re{\hbox {\rm Re}}
\def\Im{\hbox {\rm Im}}
\def\exo{\hbox {\rm exp}}
\def\diag{\hbox{\rm diag}}
\def\longvert{{\rule[-2mm]{0.1mm}{7mm}}\,}
\def\a{\alpha}
\def\dag{{}^{\dagger}}
\def\tq{{\widetilde q}}
\def\p{{}^{\prime}}
\def\W{W}
\def\N{{\cal N}}
\def\hsp{,\hspace{.7cm}}

\def\br{\nonumber}
\def\IZ{{\mathbb Z}}
\def\IR{{\mathbb R}}
\def\IC{{\mathbb C}}
\def\IQ{{\mathbb Q}}
\def\IP{{\mathbb P}}
\def \eqn#1#2{\begin{equation}#2\label{#1}\end{equation}}

\newcommand{\C}{\ensuremath{\mathbb C}}
\newcommand{\Z}{\ensuremath{\mathbb Z}}
\newcommand{\R}{\ensuremath{\mathbb R}}
\newcommand{\rp}{\ensuremath{\mathbb {RP}}}
\newcommand{\cp}{\ensuremath{\mathbb {CP}}}
\newcommand{\vac}{\ensuremath{|0\rangle}}
\newcommand{\vact}{\ensuremath{|00\rangle}                    }
\newcommand{\oc}{\ensuremath{\overline{c}}}
\newcommand{\psizero}{\psi_{0}}
\newcommand{\phizero}{\phi_{0}}
\newcommand{\hzero}{h_{0}}
\newcommand{\psiin}{\psi_{\rh}}
\newcommand{\phiin}{\phi_{\rh}}
\newcommand{\hin}{h_{\rh}}
\newcommand{\rh}{r_{h}}
\newcommand{\rb}{r_{b}}
\newcommand{\psibnd}{\psi_{0}^{b}}
\newcommand{\psibndp}{\psi_{1}^{b}}
\newcommand{\phibnd}{\phi_{0}^{b}}
\newcommand{\phibndp}{\phi_{1}^{b}}
\newcommand{\gbnd}{g_{0}^{b}}
\newcommand{\hbnd}{h_{0}^{b}}
\newcommand{\zh}{z_{h}}
\newcommand{\zb}{z_{b}}
\newcommand{\man}{\mathcal{M}}
\newcommand{\hbr}{\bar{h}}
\newcommand{\tbr}{\bar{t}}

\begin{titlepage}
\begin{flushright}
CERN-TH-2023-187 
\end{flushright}

\def\thefootnote{\fnsymbol{footnote}}

\begin{center}
{
{\bf {\large Brickwall in Rotating BTZ: A Dip-Ramp-Plateau Story} 
}
}
\end{center}

\begin{center}
Suman Das$^a$\footnote{\texttt{suman.das[at]saha.ac.in}}, Arnab Kundu$^{a,b}$\footnote{\texttt{arnab.kundu[at]saha.ac.in}}

\end{center}

\renewcommand{\thefootnote}{\arabic{footnote}}

\begin{center}

$^a$ {Theory Division, Saha Institute of Nuclear Physics, \\A CI of Homi Bhabha National Institute,
1/AF, Bidhannagar, Kolkata 700064, India.}\\

\bigskip

$^b$ Department of Theoretical Physics, CERN, CH-1211 Geneva 23, Switzerland.

\end{center}

\noindent
\begin{center} {\bf Abstract} \end{center}

In this article, building on our recent investigations and motivated by the fuzzball-paradigm, we explore normal modes of a probe massless scalar field in the rotating BTZ-geometry in an asymptotically AdS spacetime and correspondingly obtain the Spectral Form Factor (SFF) of the scalar field. In particular, we analyze the SFF obtained from the single-particle partition function. We observe that, a non-trivial Dip-Ramp-Plateau (DRP) structure, with a Ramp of slope one (within numerical precision) exists in the SFF which is obtained from the grand-canonical partition function. This behaviour is observed to remain stable close to extremality as well. However, at exact extremality, we observe a loss of the DRP-structure in the corresponding SFF. Technically, we have used two methods to obtain our results: (i) An explicit and direct numerical solution of the boundary conditions to obtain the normal modes, (ii) A WKB-approximation, which yields analytic, semi-analytic and efficient numerical solutions for the modes in various regimes. We further re-visit the non-rotating case and elucidate the effectiveness of the WKB-approximation in this case, which allows for an analytic expression of the normal modes in the regime where a level-repulsion exists. This regime corresponds to the lower end of the spectrum as a function of the scalar angular momentum, while the higher end of this spectrum tends to become flat. By analyzing the classical stress-tensor of the probe sector, we further demonstrate that the back-reaction of the scalar field grows fast as the angular momenta of the scalar modes increase in the large angular momenta regime, while the back-reaction remains controllably small in the regime where the spectrum has non-trivial level correlations. This further justifies cutting the spectrum off at a suitable value of the scalar angular momenta, beyond which the scalar back-reaction significantly modifies the background geometry.

\vspace{1.6 cm}
\vfill

\end{titlepage}

\setcounter{footnote}{0}

\section{Introduction}

Understanding and identifying the signatures of unitarity in a quantum dynamics is a non-trivial problem in physics. First, one needs a suitable diagnostic observable to both calculate and measure that can detect {\it e.g.}~the lack of unitarity. Secondly, since explicit time-dependence is generally involved for any realistic system, one needs to track the system for a very long time-scale to recover unitarity.\footnote{For example, for a system with discrete energy levels, this time-scale behaves as $t_H \sim \Delta^{-1}$, where $\Delta$ is the minimum gap in the spectrum. Clearly, this is proportional to the exponential of entropy, and is therefore extremely large for a system with many degrees of freedom.} Close to these, are ideas and notions of quantum chaos. Spectral Form Factor (SFF) and Level Spacing Distribution (LSD) provide us with a very useful diagnostic of the underlying integrable or chaotic dynamics. The extreme version of quantum chaos is generally provided by systems under the Random Matrix Theory (RMT) universality class, which displays a Dip-Ramp-Plateau (DRP) structure in the SFF, with a ramp of slope unity and a Wigner-Dyson LSD. While RMT-class systems ensure the collective existence of all these features simultaneously, it is still interesting to understand whether each of the above features can be isolated from the others. Especially to this is the existence of the slope unity ramp, which is thought to originate from the level-repulsion of far-away eigenvalues of the spectrum.

Understanding quantum aspects of gravity and black holes has been a driving force in decades of research in theoretical physics. One of the primary upshots is a potent debate about the smoothness of the event horizon of a black hole. This conflict of ideas underlies several burning issues in quantum aspects of gravity, especially unitarity\cite{Hawking:1976ra, Page:1993wv}. In particular, it has been argued,{\it e.g.}~in \cite{Mathur:2009hf, Almheiri:2012rt} that quantum gravitational effects make the event horizon non-smooth.\footnote{More specifically, it has been argued that there is a conflict between effective field theories, smooth horizons and strong sub-additivity of entanglement entropy.} Much earlier to these, 't Hooft had considered a brickwall model of a (quantum) black hole\cite{tHooft:1984kcu}, in which a Dirichlet hypersurface was placed {\it ad hoc} in front of the event horizon of a black hole.

In \cite{Das:2022evy, Das:2023ulz, Das:2023yfj} we have initiated a re-visit of the brickwall-type model, especially within the context of unitarity of the spectral form factor (SFF) of a probe (scalar) field in the given brickwall geometry. Recently, in \cite{Cotler:2016fpe}, an explicit calculation of the SFF of a quantum black hole has been carried out In a low-dimensional model and it was demonstrated to fall under the RMT-universality class. For related studies in the SYK model or in JT-gravity, see {\it e.g.}~\cite{CVJ, SSS, Altland, Jeff, Liu:2016rdi, Krishnan:2016bvg, delCampo:2017bzr, Krishnan:2017ztz, Gaikwad:2017odv, Krishnan:2017lra, Bhattacharya:2017vaz, Bhattacharya:2018nrw, Johnson:2021owr, Chen:2022hbi, Amit, Mertens, Susskind, HP}. However, at present, no analogous calculation exists in higher dimensions where gravitons are true dynamical degrees of freedom. More directly, it has been established in \cite{Cotler:2016fpe} that the SYK model and its close cousins are under the RMT universality class. Gravitational wormholes play a crucial and a rather interesting role in these studies, see {\it e.g.}~\cite{Kundu:2021nwp} for a review on certain basic aspects of wormholes from this perspective.

Although our explicit model calculations are performed in $(2+1)$-dimensions where only boundary gravitons are dynamical, we have seen strong hints of the dynamical origin of the chaotic features: the large blue shift close to the black hole geometry coming from the Rindler near-horizon geometry. Therefore, it is reasonable to expect that our observations are robust across dimensions, although an explicit check will be technically rather cumbersome at this point.

Our underlying finer-grained motivation is rooted in the fuzzball-type scenario, in which it is thought to be difficult to capture scrambling and chaotic properties of a black hole, associated with its thermal description. At this point, it is worthwhile to emphasize that the fuzzball-type scenario does not have any wormhole-type realization by construction and therefore it is potentially fruitful to compare and contrast results obtained by these two methods. Our model consists of an {\it ad hoc} Dirichlet boundary condition on a stretched horizon which is localized at a fixed radial position. This can be viewed as a toy model for a fuzzball, see {\it e.g.}~\cite{Mathur, Rychkov, KST, Avinash, superstrata, MathurReview, BenaReview, Martinec, Bena:2022ldq, Martinec:2020cml, Bena:2020yii, Warner:2019jll, Bena:2019azk, Bena:2018mpb, Tyukov:2017uig}. While, generically, the fuzzballs are characterized by a non-trivial profile function, in \cite{Das:2022evy, Das:2023ulz, Das:2023yfj} we observed several interesting and robust features with a simple Dirichlet boundary condition for a first-quantized probe scalar field and its corresponding single-particle partition function.\footnote{Note that, in \cite{Das:2023ulz}, we have explicitly explored the role of an angle-dependent Dirichlet boundary and found that, choosing a Gaussian random distribution with an order one variance produces a Wigner-Dyson like level-spacing distribution.} These are:
\begin{itemize}
\item {The spectral form factor displays a robust dip-ramp plateau structure, with ramp of slope unity.}\footnote{These features are visible in a log-log plot.}
\item {No averaging is needed to observe the linear ramp.}
\item {The single-particle spectrum displays level-repulsion.}
\end{itemize}

Motivated by these observations, in this article, we explore a similar question in the rotating BTZ geometry. There are multiple motivations of our present work: (i) First, rotating BTZ geometry is a natural and interesting generalization of our earlier works. (ii) Secondly, rotating BTZ geometry allows for the possibility of considering a grand-canonical ensemble for the probe scalar field and the corresponding analytically continued SFF. We will observe that this plays a crucial role and the above observations will hold provided we consider the SFF obtained from the grand partition function. (iii) Thirdly, rotating BTZ allows for an extremal limit which has several unique features associated to the black hole dynamics. We will observe, among other things, that the near-extremal physics is qualitatively different from the physics at exact extremality. At exact extremality both level-repulsion and the DRP-structure of the corresponding SFF disappear. This further demonstrates that the near-horizon Rindler structure along with a compact direction to support a non-vanishing angular momentum are the crucial ingredients for this behaviour. (iv) Fourth, at a technical level, the salient features related to quantum chaos in \cite{Das:2022evy, Das:2023ulz} originate from the behaviour of the normal modes as a function of the non-vanishing angular momenta of the scalar field. It is thus interesting to understand the role of a global non-vanishing angular momentum in this framework. For rotating BTZ, the black hole angular momentum provides us with this additional scale. It is intriguing that at the maximal allowed value of the BTZ angular momentum, the normal modes become a linear function of the angular momenta of the scalar sector and therefore reduce to a harmonic oscillator like behaviour.\footnote{Note that, even for the harmonic oscillator the level spacing distribution does display an extreme level repulsion, since all energy levels are equispaced. The corresponding SFF, however, does not display any chaotic behaviour.}

Before we end this section, let us summarize the key observations of this work:
\begin{itemize}
\item{For a generic non-vanishing angular momentum of the BTZ-background, the spectral form factor of the probe scalar field displays a robust dip-ramp plateau structure, with ramp of slope unity.}\footnote{These features are, once again, visible in a log-log plot.} It is perhaps worth emphasizing that we need not carry out any averaging to observe these features. However, an averaging may be done to sharpen the features as a convenience.
\item{The above observations are done from an SFF which is obtained by analytically continuing the grand-canonical partition function.  It is noteworthy that the SFF obtained from the canonical partition function does not exhibit a stable dip-ramp-plateau structure. Instead, it transitions smoothly from chaotic RMT type to simple harmonic oscillator (SHO) type, as the black hole's angular momentum increases towards extremality. } 
\item{The grand-canonical partition function can be equivalently viewed as emerging from restricting the partition sum on positive modes only. These modes display the qualitatively similar level-correlation that we have seen in our earlier works as well.}
\item{Close to extremality, the DRP features remain unaffected. To the extent we could check within our numerical constraints.}
\item{At exact extremality, there is a qualitatively different physics. The SFF now appears to resemble an integrable system and the normal modes seem to be linear and therefore harmonic oscillator like. A clear demarcation between the close-to-extremality and exact-extremality is the precise nature of red-shift produced by the black hole geometry. While, at exact extremality, the surface gravity vanishes identically; close to extremality it is arbitrarily small but non-vanishing. Equivalently, the former corresponds to physics at exact $T=0$ temperature, whereas the latter corresponds to a small non-vanishing temperature. }
\item{All our statements above are made for the single-particle Spectral Form Factor.}
\end{itemize}

This article is divided into the following parts: In the next section, we begin with our set-up describing a probe Klein-Gordon field in the rotating BTZ geometry. We first review, using the previously unused WKB-approximation, the non-rotating case that was reported in our earlier work. This approximation allows us for certain analytic regimes as well as faster and more efficient numerical solutions. Subsequently we discuss the rotating BTZ case. We use numerical methods, analytical and semi-analytical (WKB) methods in exploring this case in general. We then analyze the exact extremal case separately, using the WKB-approximation. We then conclude with several open directions as well as comments on our current works in progress in the Discussion section. We have relegated several technical details to five appendices.

\section{Probe Scalar in Rotating Black Hole Background}

Let's consider rotating BTZ metric in $2+1$ dimensions,
\begin{equation}\label{metric}
    ds^2=g_{tt}dt^2+g_{rr}dr^2+2 g_{t\psi} dt d\psi+g_{\psi \psi} d\psi^2 ,
\end{equation}
where,
\begin{eqnarray}
g_{tt}=M-\frac{r^2}{l^2}, \ \ g_{t\psi}=-\frac{J}{2}, \ \ g_{\psi \psi}=r^2  \ \ {\rm and} \ \ g_{rr}=\left(-M+\frac{J^2}{4r^2}+\frac{r^2}{l^2}\right)^{-1} . \label{metcomp}
\end{eqnarray}
Here we have set $c=1$ throughout the paper. $M$ and $J$ are respectively mass and angular momentum of the black hole. The inner and outer (event horizon) horizons are defined by,
\begin{equation}\label{horizon}
   r_{\pm}^2=\frac{M l^2}{2} \left(1\pm \sqrt{1-\frac{J^2}{M^2 l^2}}\right) .
\end{equation}
We want to quantize a massless probe scalar field in this background i.e. we want solve the following Klein-Gordon equation,
\begin{equation}\label{eom1}
    \Box \Phi\equiv \frac{1}{\sqrt{|g|}}\partial_{\mu}\left(\sqrt{|g|}\partial^{\mu}\Phi\right)=\mu^2 \Phi^2.
\end{equation}
As the metric is invariant under the translation of $t$ and $\psi$, we will use the ansatz, $ \Phi = \sum_{\omega, m} e^{-i\omega t}  e^{i m\psi} \phi_{\omega, m}(r)$ and with this ansatz radial part of \eqref{eom1} satisfies,
\begin{equation}\label{radial}
    \Bigg(g_{rr}\left( \omega-\frac{J}{2r^2}m \right)^2-\frac{\mu^2}{r^2}+\frac{1}{r}\frac{d}{dr}\left( \frac{r}{g_{rr}}\frac{d}{dr} \right)\Bigg)\phi(r)=0.
\end{equation}
Here we have suppressed the subscript of $\phi(r)$. To write \eqref{radial} in more familiar form, we use a new radial coordinate $z=\frac{r^2-r_+^2}{r^2-r_-^2}$ and introduce a new radial function $F(z)=z^{i\alpha}(1-z)^{-\beta} \phi(r)$, where
\begin{eqnarray}
    \alpha=\frac{l^2 r_+}{2(r_+^2-r_-^2)}(\omega-\Omega_H m), \ \ \beta=\frac{1}{2}(1-\sqrt{1+\mu^2}).
\end{eqnarray}
Here $\Omega_H=J/2 r_+^2$ is the angular velocity at the horizon. In this new coordinate, $z\rightarrow1$ is the boundary and $z\rightarrow0$ corresponds to outer horizon. With this changes \eqref{radial} simplifies to,
\begin{equation}\label{hyper}
    z(1-z)\frac{d^2 F(z)}{dz^2}+(c-(1+a+b)z)\frac{dF(z)}{dz}-ab F(z)=0,
\end{equation}
with
\begin{eqnarray}
    a=\beta-i \frac{l^2}{2(r_++r_-)}\left(\omega+\frac{m}{l}\right), \ \ b=\beta-i \frac{l^2}{2(r_++r_-)}\left(\omega-\frac{m}{l}\right), \ \ c=1- 2i \alpha ,
\end{eqnarray}
solution is given by ,
\begin{align}\label{sol1}
    \phi(z)=z^{-i \alpha}(1-z)^{\beta} ( C_1 & \, {}_2F_1(a, b; 1+a+b-c; 1-z)+ \nonumber \\
    & C_2 (1-z)^{c-a-b} \, {}_2F_1(c-a, c-b; 1+c-a-b; 1-z))
\end{align}
Near boundary behaviour of \eqref{sol1} is
\begin{equation}
    \phi_{\text{bdry}}(z)\approx z^{-i \alpha} \left( C_1 (1-z)^{\frac{1}{2}(1-\sqrt{1+\mu^2})} +  C_2 (1-z)^{\frac{1}{2}(1+\sqrt{1+\mu^2})} \right),
\end{equation}
where the first term is non-normalizable and second one is normalizable. So normalizable condition at boundary implies $C_1=0$, i.e,
\begin{equation}
    \begin{split}
        \phi(z) & \sim z^{-i \alpha} (1-z)^{\beta} \, (1-z)^{c-a-b} \, {}_2F_1(c-a, c-b; 1+c-a-b; 1-z) \\
        & = z^{-i \alpha}(1-z)^{\beta} \left(P \,  {}_2F_1(a, b; c; z) +Q z^{1-c}  {}_2F_1(a-c+1, b-c+1; 2-c; z)   \right)\\
        & = (1-z)^{\beta} \left(P \, z^{-i \alpha} \, {}_2F_1(a, b; c; z) +Q \, z^{i \alpha}  {}_2F_1(a-c+1, b-c+1; 2-c; z)   \right)
    \end{split}
\end{equation}
where,
\begin{equation}
    P=\frac{\Gamma(1-c) \Gamma(c-a-b+1)}{\Gamma(1-a)\Gamma(1-b)}, \hspace{1cm} Q= \frac{\Gamma(c-1)\Gamma(c-a-b+1)}{\Gamma(c-a) \Gamma(c-b)}
\end{equation}
Dirichlet boundary condition that field vanishes at some $z=z_0$ implies:
\begin{equation}
    \frac{Q}{P}=-z_0^{-2 i \alpha} \frac{ {}_2F_1(a, b; c; z_0)}{{}_2F_1(a-c+1, b-c+1; 2-c; z_0)} \ . \label{Dbchor}
\end{equation}
This is the quantization condition which gives rise to normal modes. When the position of the stretched horizon is very close to the event horizon ($z_0\rightarrow 0$), we can approximate the ratio as $-z_0^{-2 i \alpha}$ and then quantization condition becomes,
\begin{eqnarray}\label{quant1}
     \frac{Q}{P}= \frac{\Gamma(c-1)}{\Gamma(c-a) \Gamma(c-b)} \frac{\Gamma(1-a)\Gamma(1-b)}{\Gamma(1-c)}=-z_0^{-2i \alpha} \nonumber \\
    \Rightarrow -z_0^{2 i \alpha}  \frac{\Gamma(c-1)}{\Gamma(c-a) \Gamma(c-b)} \frac{\Gamma(1-a)\Gamma(1-b)}{\Gamma(1-c)}=1  \nonumber \\
    \Rightarrow  \text{Arg} \left( \frac{\Gamma(c-1)}{\Gamma(c-a) \Gamma(c-b)}  \right)+ \alpha \log z_0=-\left(n+\frac{1}{2} \right)\pi.
\end{eqnarray}
Let $\epsilon=r_0-r_+$ denotes the separation between horizon and stretched horizon. Then \eqref{quant1} can be written as the following,
\begin{equation}\label{quantizn_eqn}
    \alpha \log \left(  \frac{2 \, \epsilon \, r_+}{r_+^2-r_-^2} \right)+ \text{Arg} \left( \frac{\Gamma(c-1)}{\Gamma(c-a) \Gamma(c-b)}  \right)=-\left(n+\frac{1}{2} \right)\pi,  \hspace{1cm}\text{where}, n\in \mathbf{Z}.
\end{equation}
We have solved this equation in Mathematica which gives us normal modes $\omega(n,m)$ as a function of principal quantum number $n$ and rotational quantum number $m$. In the subsequent sections we will consider those modes for which $\omega(n,m)-\Omega_H m>0$, i.e. when partition sum is well defined. Before that, let us revisit the non-rotating BTZ geometry first.

\subsection{Non-rotating BTZ: $J=0$ case}

\begin{figure}[H]
\begin{subfigure}{0.47\textwidth}
    \centering
    \includegraphics[width=\textwidth]{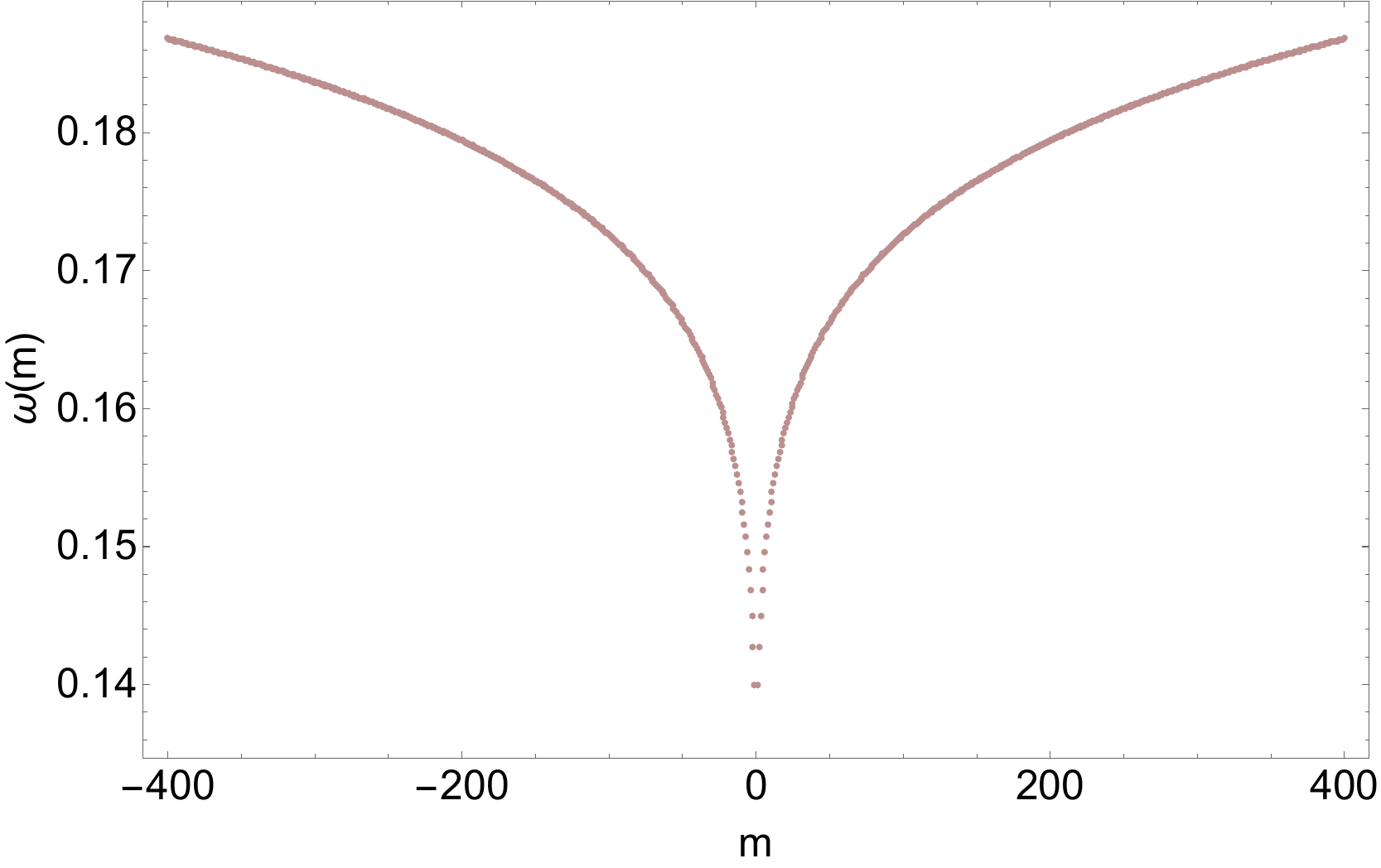}
    \end{subfigure}
    \hfill
    \begin{subfigure}{0.47\textwidth}
    \includegraphics[width=\textwidth]{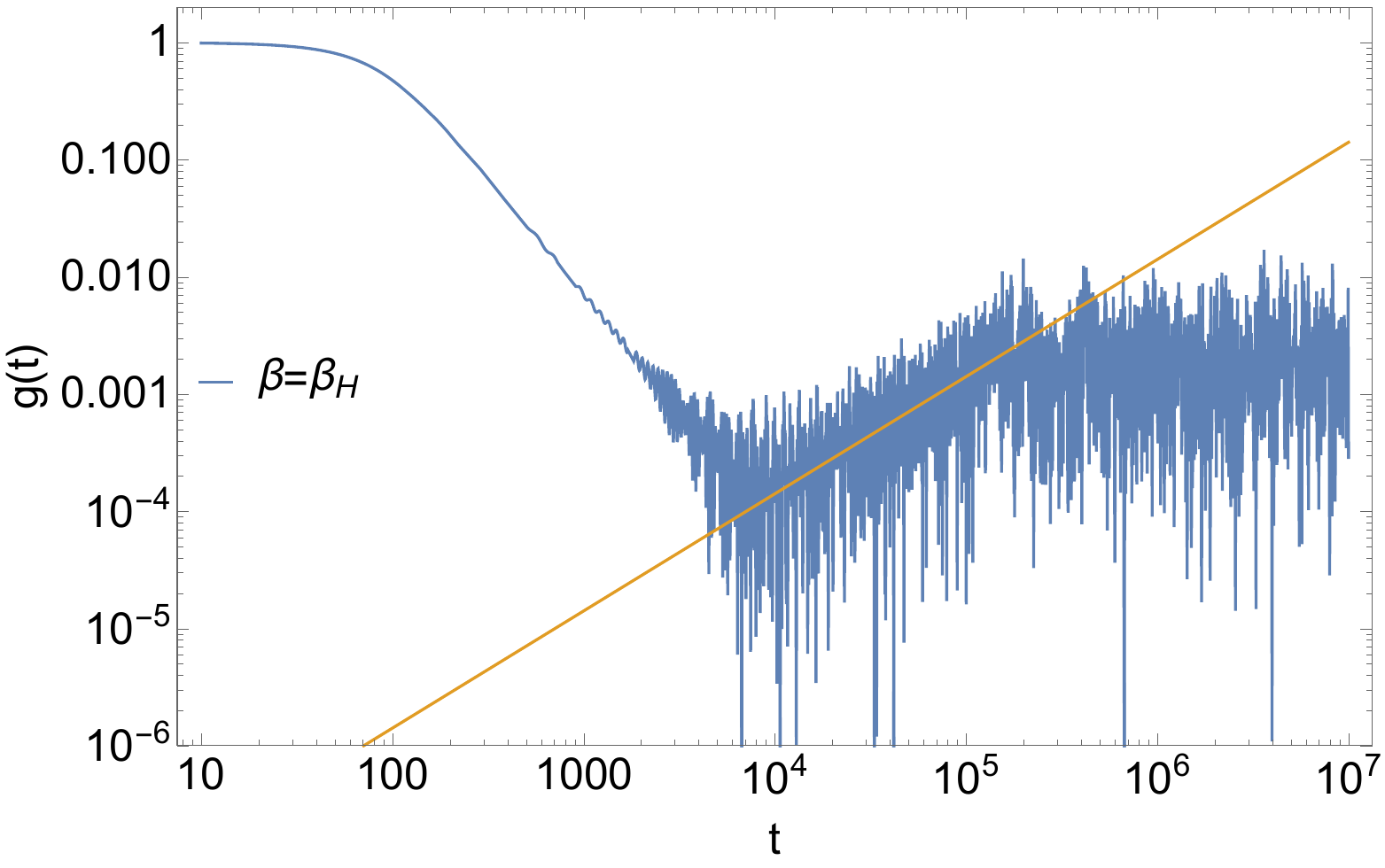}
    \end{subfigure}
    \caption{Spectrum (left) and SFF (right) for static BTZ i.e. $J=0$ case along $m$-direction. Here $n=0$, $m_{cut}=400$ and $z_0=10^{-20}$ and $\beta=\beta_H$. The yellow line has slope one.}
    \label{J_0_spectrum}
\end{figure}

Let's consider $J=0$ case first i.e. static BTZ black hole which we have already studied in \cite{Das:2022evy}. As \eqref{quantizn_eqn} is symmetric under $m\rightarrow -m$, roots i.e. normal modes also preserve the symmetry as shown in Figure \ref{J_0_spectrum} (left). Corresponding spectral form factor is shown in the right of the Figure \ref{J_0_spectrum}.
%

\subsubsection{Some Analytical Regimes \& Estimates}

\begin{figure}[H]
    \centering
    \includegraphics[width=.60\textwidth]{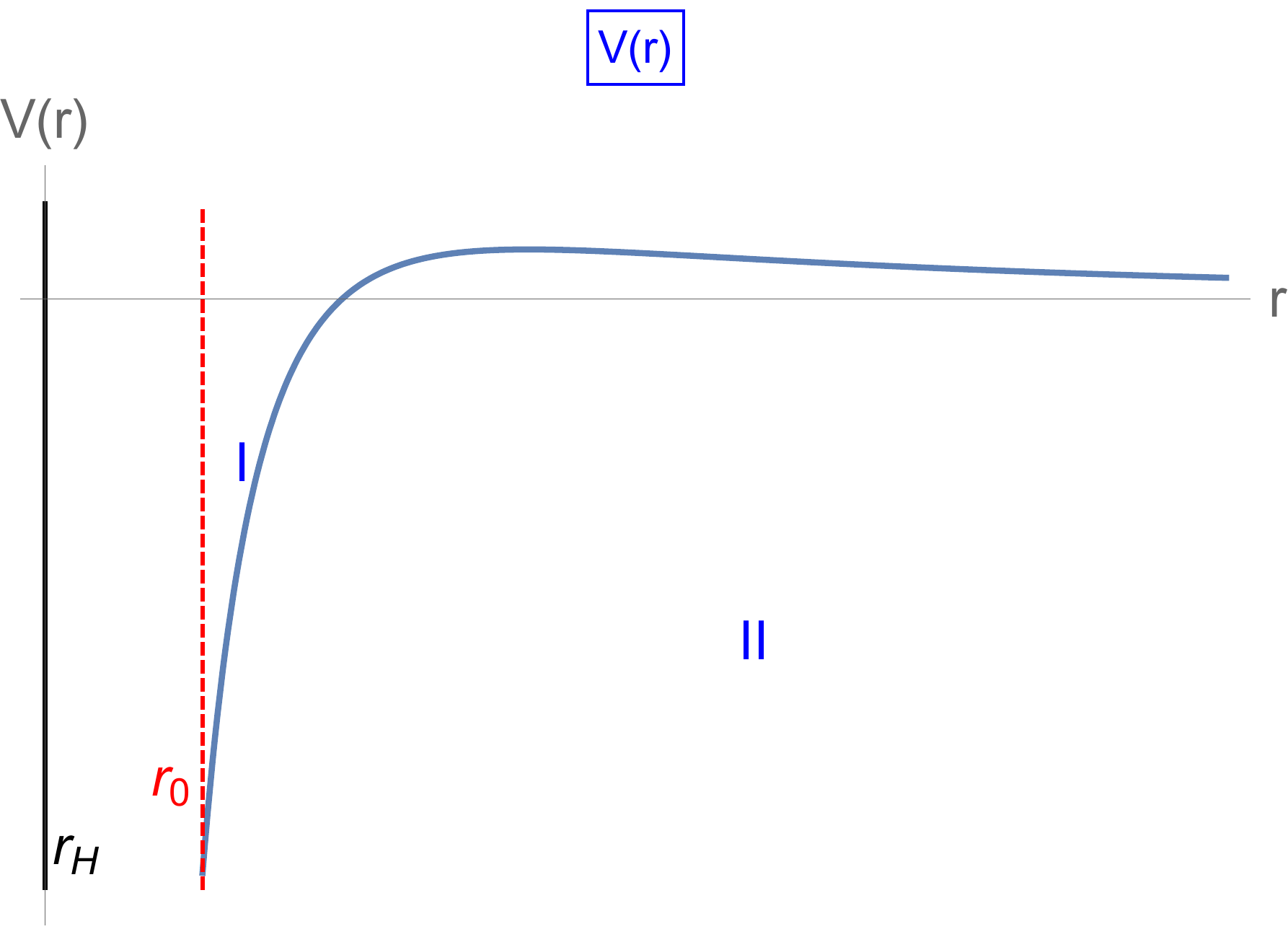}
   \caption{The Schr\"{o}dinger potential corresponding to the Klein-Gordon equation in non-rotating BTZ-background. We have chosen $m=1.6$ and $\omega=1$ for this particular case. The shape of the potential shown in this figure is, however, generic. The robust qualitative feature is that there exists two regions $I$ and $II$, where $V(r) <0$ and $V(r)>0$, respectively. These two regions are separated by a classical turning point, denoted by $r_c$ where $V(r_c)=0$. Also, $r_H$ and $r_0$ correspond to the location of the event horizon and the cut-off surface, respectively.}
    \label{wkbpot_noJ}
\end{figure}

In this section, we will use the WKB approximation to explore analytical regimes of the spectrum. Furthermore, as we will explicitly demonstrate, the WKB-approximation is also efficient in numerically obtaining the normal modes. Some of the key details are provided in appendix \ref{KGSch} and we will use them in this section. First of all, note that the Klein-Gordon equation, given explicitly in the $r$ coordinate of \cite{Das:2022evy}, can be written as:
\begin{eqnarray}
\left( r^2 -1\right)^2 \phi''(r) + 2r (r^2-1) \phi'(r) + \left(\omega^2 - (r^2 -1) \left( \frac{1}{r^2} \left( m^2 + \frac{1}{4} \right) + \frac{3}{4} \right)  \right) \phi(r) = 0 \ .
\end{eqnarray}
This equation can now be written in the form of a Schr\"{o}dinger equation:
\begin{eqnarray}
\frac{d^2 \Psi}{dr^2} - V(r) \Psi(r) = 0 \ .
\end{eqnarray}
The explicit form of the potential can be obtained from equation (\ref{schpot}), which we will not explicitly present here. Instead, we will refer the Reader to a generic form of the potential in {\it e.g.}~figure \ref{wkbpot_noJ}.

It is clear from the pictorial representation of the potential that there are three regions where the WKB approximation yields analytic solutions, at least in terms of integrals. The full solution of the wavefunction is given by (in what follows, we closely follow the notations of \cite{Bena:2019azk})
\begin{eqnarray}
\Psi(r) & = & \frac{1}{|V(r)|^{\frac{1}{4}}} \left[ D_+^I {\rm exp} \left( i \int_r^{r_c} |V(r)|^{\frac{1}{2}} dr\right) + D_-^I {\rm exp} \left( - i \int_r^{r_c} |V(r)|^{\frac{1}{2}} dr\right) \right] \ , r < r_c \ , \\
& = & d_+^I {\rm Bi} \left( V'(r_c)^{\frac{1}{3}} (r-r_c) \right) + d_-^I {\rm Ai} \left( V'(r_c)^{\frac{1}{3}} (r-r_c) \right) \ , r \sim r_c \ , \\
& = &  \frac{1}{|V(r)|^{\frac{1}{4}}} \left[ D_+^{II} {\rm exp} \left( \int_{r_c}^{r} |V(r)|^{\frac{1}{2}} dr\right) + D_-^{II} {\rm exp} \left( -  \int_{r_c}^{r} |V(r)|^{\frac{1}{2}} dr\right) \right] \ , r > r_c \ .
\end{eqnarray}
As obtained in \cite{Bena:2019azk}, the corresponding WKB connection formulae relate the coefficients $\{d_+^I, d_-^I\}$ with $\{D_+^I, D_-^{I}\}$ and $\{d_+^I, d_-^I\}$ with $\{D_+^{II}, D_-^{II}\}$:
\begin{eqnarray}
&& \begin{bmatrix} 
	d_+^I  \\
	d_-^I \\
	\end{bmatrix} = e^{- i \frac{\pi}{4}}\sqrt{\pi} V'(r_c)^{- \frac{1}{6}} \begin{bmatrix} 
	1 & i  \\
	i & 1 \\
	\end{bmatrix} \begin{bmatrix} 
	D_+^I  \\
	D_-^I \\
	\end{bmatrix}  \ , \\
&&	 \begin{bmatrix} 
	d_+^I  \\
	d_-^I \\
	\end{bmatrix} = \sqrt{\pi} V'(r_c)^{- \frac{1}{6}} \begin{bmatrix} 
	1 & 0  \\
	0 & 2 \\
	\end{bmatrix} \begin{bmatrix} 
	D_+^{II}  \\
	D_-^{II} \\
	\end{bmatrix}  \ , 	
\end{eqnarray}
which implies
\begin{eqnarray}
 \begin{bmatrix} 
	D_+^I  \\
	D_-^I \\
	\end{bmatrix} = \sqrt{\pi} V'(r_c)^{- \frac{1}{6}} \begin{bmatrix} 
	\frac{	1}{2}e^{\frac{i\pi}{4}} & e^{-\frac{i\pi}{4}}   \\
	\frac{	1}{2}e^{- \frac{i\pi}{4}} &  e^{i \frac{i\pi}{4}} \\
	\end{bmatrix} \begin{bmatrix} 
	D_+^{II}  \\
	D_-^{II} \\
	\end{bmatrix}  \ .
\end{eqnarray}

By observation, in the regime $r>r_c$, normalizability of the solution as $r \to \infty$ requires $D_+^{II}=0$, and hence $d_{+}^I=0$ using the connection formulae. This also fixes the other constants: $d_-^I = \sqrt{\pi} V'(r_c)^{-1/6} D_-^{II}$, $D_+^I = e^{-i\pi/4} D_-^{II}$ and $D_-^{I}= e^{i\pi/4} D_-^{II}$. Hence, the explicit solution is:
\begin{eqnarray}
\Psi(r) & = & \frac{2 D_-^{II} }{|V(r)|^{\frac{1}{4}}}  \cos \left(  \int_r^{r_c} |V(r)|^{\frac{1}{2}} dr - \frac{\pi}{4} \right)   \ , \quad r < r_c \ , \\
& = &  D_-^{II} 2 \sqrt{\pi} V'(r_c)^{-\frac{1}{6}} {\rm Ai} \left( V'(r_c)^{\frac{1}{3}} (r-r_c) \right) \ , \quad r \sim r_c \ , \\
& = &  \frac{1}{|V(r)|^{\frac{1}{4}}}  D_-^{II} {\rm exp} \left( -  \int_{r_c}^{r} |V(r)|^{\frac{1}{2}} dr\right)  \ , \quad r > r_c \ .
\end{eqnarray}
The boundary condition $\Psi(r_0)=0$ now yields:
\begin{eqnarray}
\cos \left(  \int_{r_0}^{r_c} |V(r)|^{\frac{1}{2}} dr - \frac{\pi}{4} \right) = 0 \quad \implies \int_{r_0}^{r_c} |V(r)|^{\frac{1}{2}} dr - \frac{\pi}{4} = \frac{\pi}{2} + 2 n \pi \ , n\in {\mathbb Z} \ . \label{wkbrule}
\end{eqnarray}

To obtain the spectrum, let us collect some further explicit formulae. The potential, $V(r)$ is given by
\begin{eqnarray} \label{pot_noJ}
&& V(r) = \frac{r^2 \left(4 m^2-4 \omega ^2-6\right)-4 m^2+3 r^4-1}{4 r^2
   \left(r^2-1\right)^2} \ , \quad r_0 \le r \le r_c \ , \\
&& r_c^2 = \frac{1}{3} \left(-2 m^2+2 \sqrt{m^4-2 m^2 \omega ^2+\omega ^4+3
   \omega ^2+3}+2 \omega ^2+3\right) \ .   
\end{eqnarray}
Although $V(r_c)=0$ admits four distinct solutions, only the above is real and positive and hence we discard the rest. Fortunately, with the above potential the WKB-integral in (\ref{wkbrule}) can be performed analytically to yield:
\begin{eqnarray}
&& \int_{r_0}^{r_c} |V(r)|^{\frac{1}{2}} dr = \frac{1}{8} \left(-2 \sqrt{3} \tan ^{-1}(a_1)+2 \sqrt{\omega^2+1} \log \left(\frac{b_1+1}{b_1 - 1}\right) - \sqrt{4 m^2+1} \log \left(\frac{c_1+1}{c_1-1}\right)-\sqrt{3} \pi \right) \nonumber\\
&& a_1=\frac{-2 m^2-3 r_0^2+2 \omega ^2+3}{\sqrt{3} \sqrt{-4 m^2 \left(r_0^2-1\right)-3 r_0^4+r^2 \left(4 \omega^2 + 6 \right)+1}} \ , \\
&& b_1 = \frac{m^2 \left(-\left(r_0^2 - 1\right) \right) + \left(r_0^2+1\right) \omega ^2+2}{\sqrt{\omega ^2+1} \sqrt{-4 m^2 \left(r_0^2 - 1\right)-3 r_0^4 + r_0^2 \left(4 \omega^2+6\right)+1}} \ , \\
&& c_1 = \frac{-2 m^2 \left(r_0^2-2\right) + r_0^2 \left(2 \omega^2+3\right)+1}{\sqrt{4 m^2+1} \sqrt{-4 m^2 \left(r_0^2 - 1\right) - 3r_0^4 + r_0^2 \left(4 \omega^2+6\right)+1}} \ .       
\end{eqnarray}
While the above expressions are analytical, it is still difficult to invert them and obtain an analytic expression of $\omega(m)$. It is, nonetheless, possible to numerically solve the WKB-equation, obtain the normal modes and compare them with the normal modes that we have already obtained by solving the boundary conditions directly. It is also possible to obtain more intuition on the spectrum, by looking at specific regimes of the WKB-formulae above.

\begin{figure}[H]
    \centering
    \includegraphics[width=.60\textwidth]{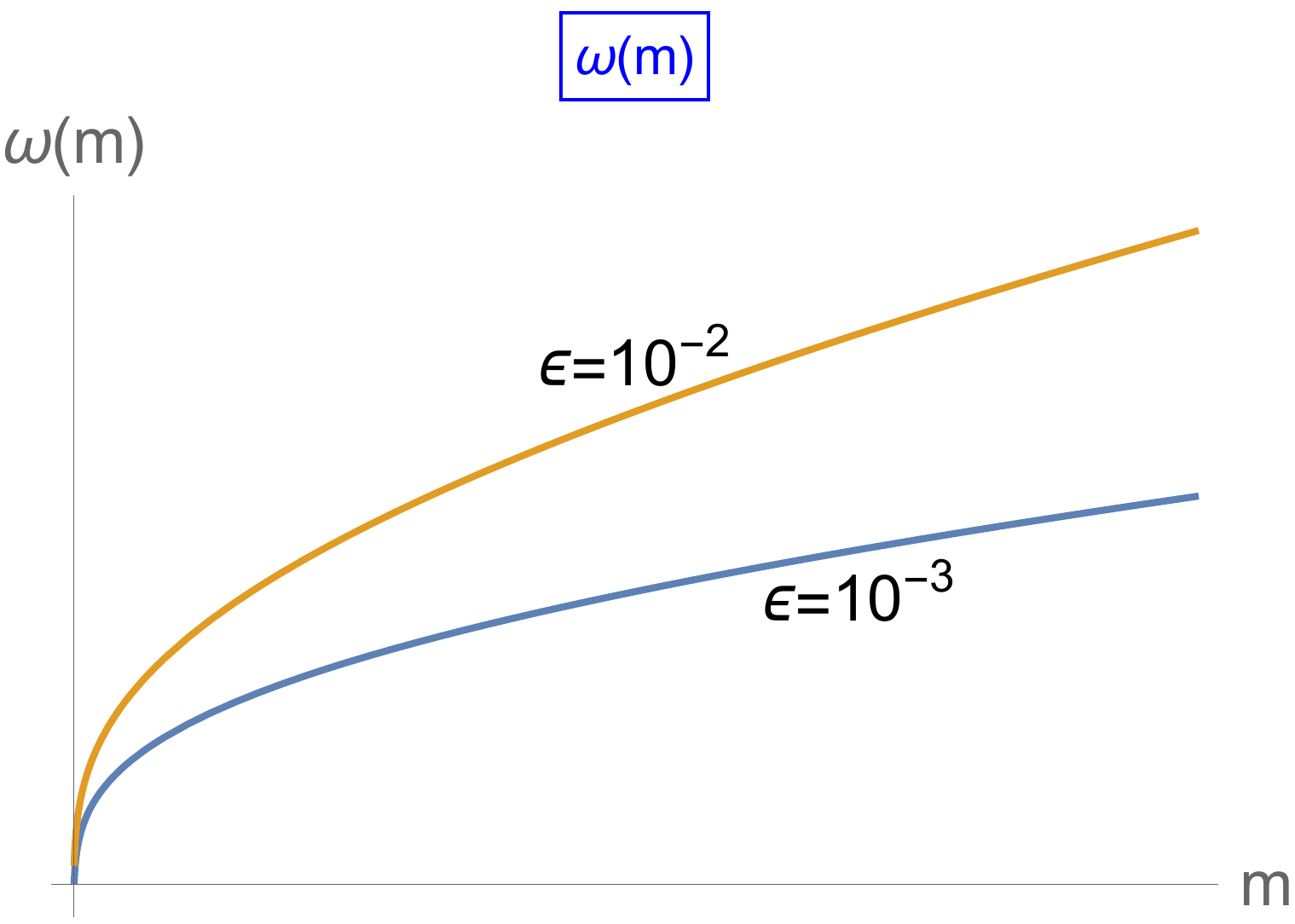}
   \caption{A pictorial representation of the analytic formula in (\ref{omegaexact}), for given choices for $\epsilon$ which has been explicitly shown in the figure above. In both cases, we have set $n=1$. Evidently, the lower ends of the spectrum coincide rather well. It is further evident from the trend of these curves that by increasing $\epsilon$, one decreases the curvature of the $\omega(m)$ function. It is thus conceivable that for large enough $\epsilon$, $\omega(m)$ will become a linear function.}
    \label{wkbeps_omega}
\end{figure}

\begin{figure}[H]
    \centering
    \includegraphics[width=.60\textwidth]{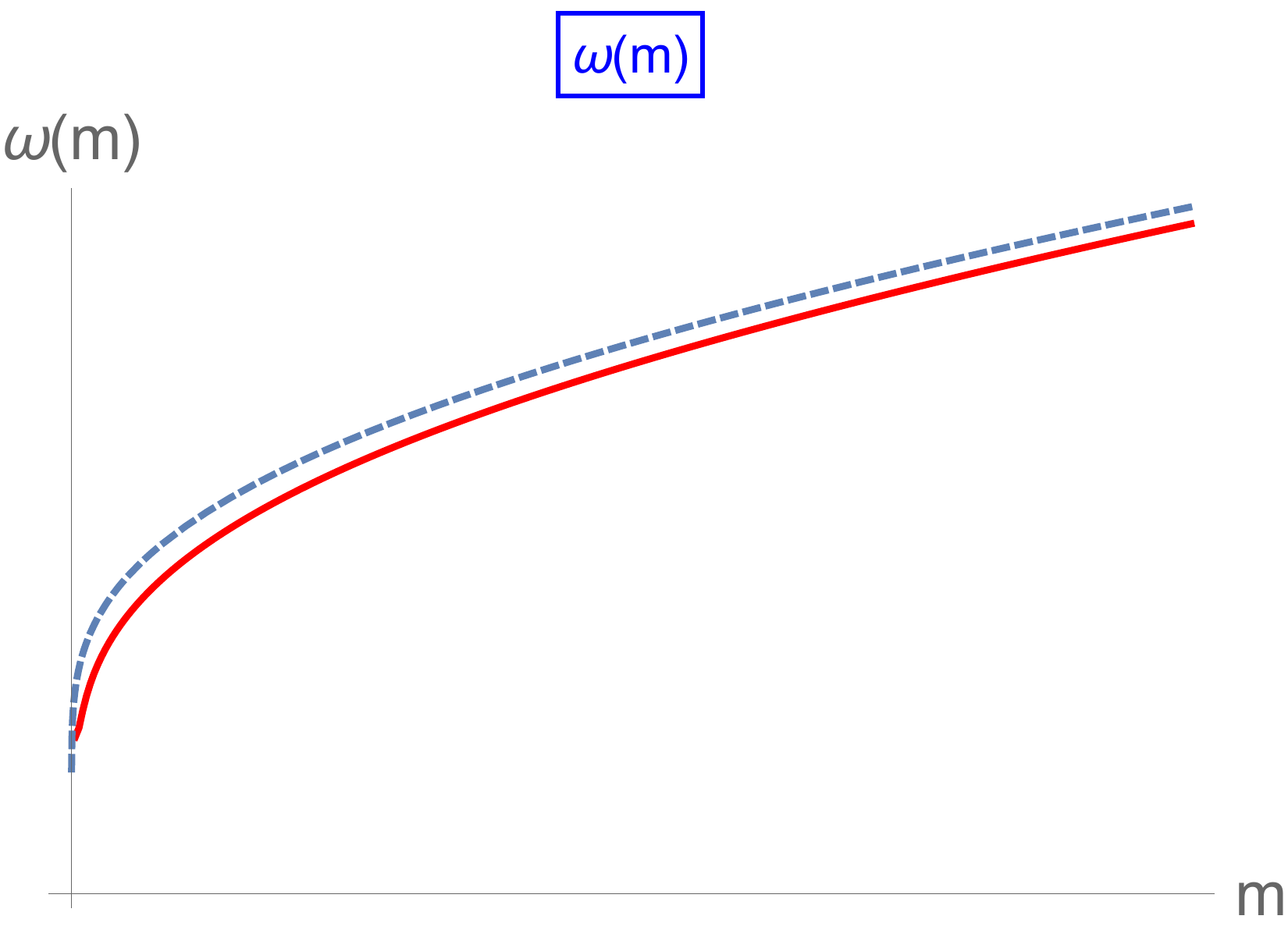}
   \caption{A pictorial representation comparing the analytic formula in (\ref{omegaexact}), for given choices for $\epsilon$ with numerical solutions obtained by the leading term in the $\epsilon\to 0$ limit. We have set $\epsilon = 10^{-5}$ and $n=1$ in the above figure. The blue dashed curve represents the formula in (\ref{omegaexact}) while the red solid curve represents numerical solutions. It is clear that (\ref{omegaexact}) provides a reliable approximation to the normal modes.}
    \label{wkb_omega_exactnumeric}
\end{figure}

\begin{figure}[H]
    \centering
    \includegraphics[width=.60\textwidth]{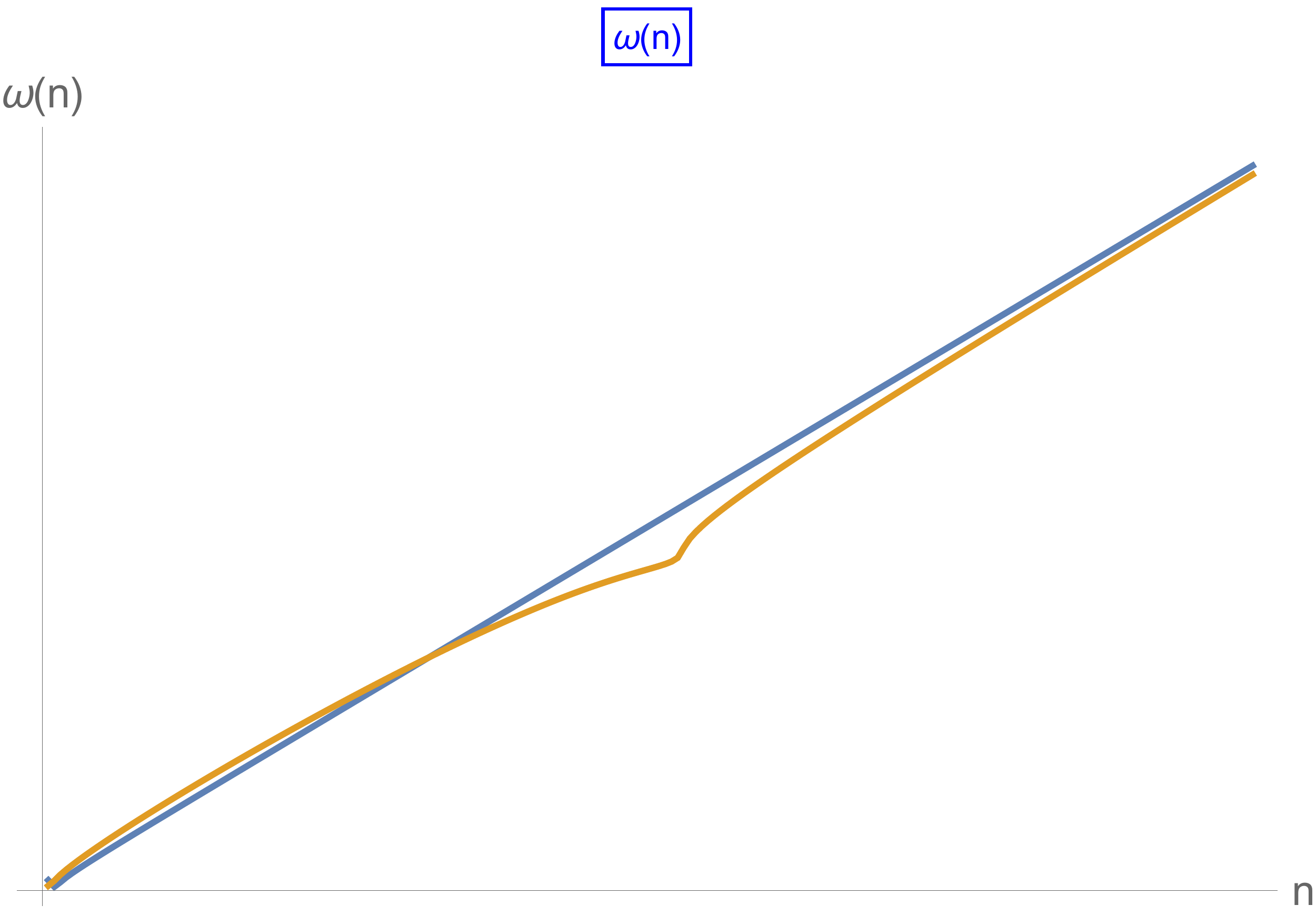}
   \caption{A pictorial representation of the dependence of normal modes with $n$, for a fixed value of the angular momentum $m$. In this plot, the blue and the yellow curves correspond to $m=1$ and $m=30$, respectively. It is evident that $\omega(n)$ closely resembles a linear function, and this dependence is rather robust against the choice of $m$. }
    \label{wkb_omega_n}
\end{figure}

We will begin with the second case. It is particularly simple to consider the cut-off surface $r_0 \to r_H$: $r_0 = r_H + \epsilon$ and perform an $\epsilon$-expansion of the above integrals. At the leading order, we obtain:
\begin{eqnarray}
&& \frac{3\pi}{4}+ 2 n \pi  =  \frac{1}{8} \left(-\sqrt{4 m^2+1} \log \left(\frac{\sqrt{4 m^2+1} \sqrt{\omega ^2+1}+m^2+\omega ^2+2}{-\sqrt{4 m^2+1} \sqrt{\omega ^2+1}+m^2+\omega ^2+2}\right) \right. \nonumber\\
&& -  \left. 2 \sqrt{3} \tan^{-1}\left(\frac{\omega ^2-m^2}{\sqrt{3} \sqrt{\omega^2+1}}\right)+2 \sqrt{\omega ^2+1} \log \left(\frac{4\left(\omega ^2+1\right)^2}{\epsilon ^2 \left(m^4-2 m^2 \omega^2+\omega ^4+3 \omega ^2+3\right)}\right)-\sqrt{3} \pi \right) \ . \nonumber\\
\end{eqnarray}
The equation above is still somewhat unwieldy to invert. We can further assume that $\omega \gg 1$ as well as $m\gg 1$ such that $\omega (m) \ll m$.\footnote{This essentially implies that $\omega(m)$ is a slowly varying function of $m$, {\it i.e.}~a logarithmic function.} In this limit, we obtain:
\begin{eqnarray}
\frac{3\pi}{4}+ 2 n \pi  = \frac{1}{4} \left(\omega  \log \left(\frac{4 \omega ^4}{m^4
   \epsilon ^2}\right)-\sqrt{3} \tan ^{-1}\left(\frac{\sqrt{3}
   \omega }{m^2}\right)\right) \ . 
\end{eqnarray}
In the $\epsilon \to 0$ limit, the first term above dominates and we obtain:
\begin{eqnarray}
\omega(m) = \frac{\pi  (8 n+3)}{4 W\left(\frac{8 \sqrt{2} \pi  n+3 \sqrt{2} \pi }{4 m \sqrt{\epsilon }}\right)} \ , \label{omegaexact}
\end{eqnarray}
where $W$ is the product log function.\footnote{The Product Log function $W(z)$ yields the principal solution for $x$ satisfying: $z = x e^x$. } The above approximation breaks down roughly at $\omega \approx m^{2/3} \epsilon^{4/3}$, which implies that we can use the approximate result for large enough angular momenta, as long as $m_{\rm max} \sim \epsilon^{-2/3}$, for a given $\epsilon$. Tuning $\epsilon$ appropriately, it is thus possible to have an access to the high-end of the $m$-spectrum for the function $\omega(m)$. Note further, that both the explicit solution in (\ref{omegaexact}) as well as the $\omega \sim m^{2/3}$ dependence satisfy the criterion $\omega(m) \ll m$, for large enough $m$. Some of these features are demonstrated in figure~\ref{wkbeps_omega}.
Alternatively, one can obtain straightforward, albeit somewhat brute force, numerical solutions of the WKB-equations. A representative such figure has been demonstrated in figures~\ref{wkb_omega_exactnumeric} and \ref{wkb_omega_n}. 
Note that the WKB approximations yields rather nice results compared to the numerical roots that we have earlier found out by solving directly the boundary conditions. It is noteworthy that although both require a numerical solution in the generic regime, the WKB-equations are substantially simpler to work with numerically.\footnote{Just as a comparison, using Mathematica, the time-scale it roughly takes to find ${\cal O}(10^3)$ roots of the Dirichlet boundary conditions in (\ref{quantizn_eqn}) is larger than finding ${\cal O}(10^4)$ numerical solutions of the WKB-equations in (\ref{wkbrule}).}

\subsection{Non-vanishing rotation: $J \neq 0$}

\begin{figure}[H]
\begin{subfigure}{0.47\textwidth}
    \centering
    \includegraphics[width=\textwidth]{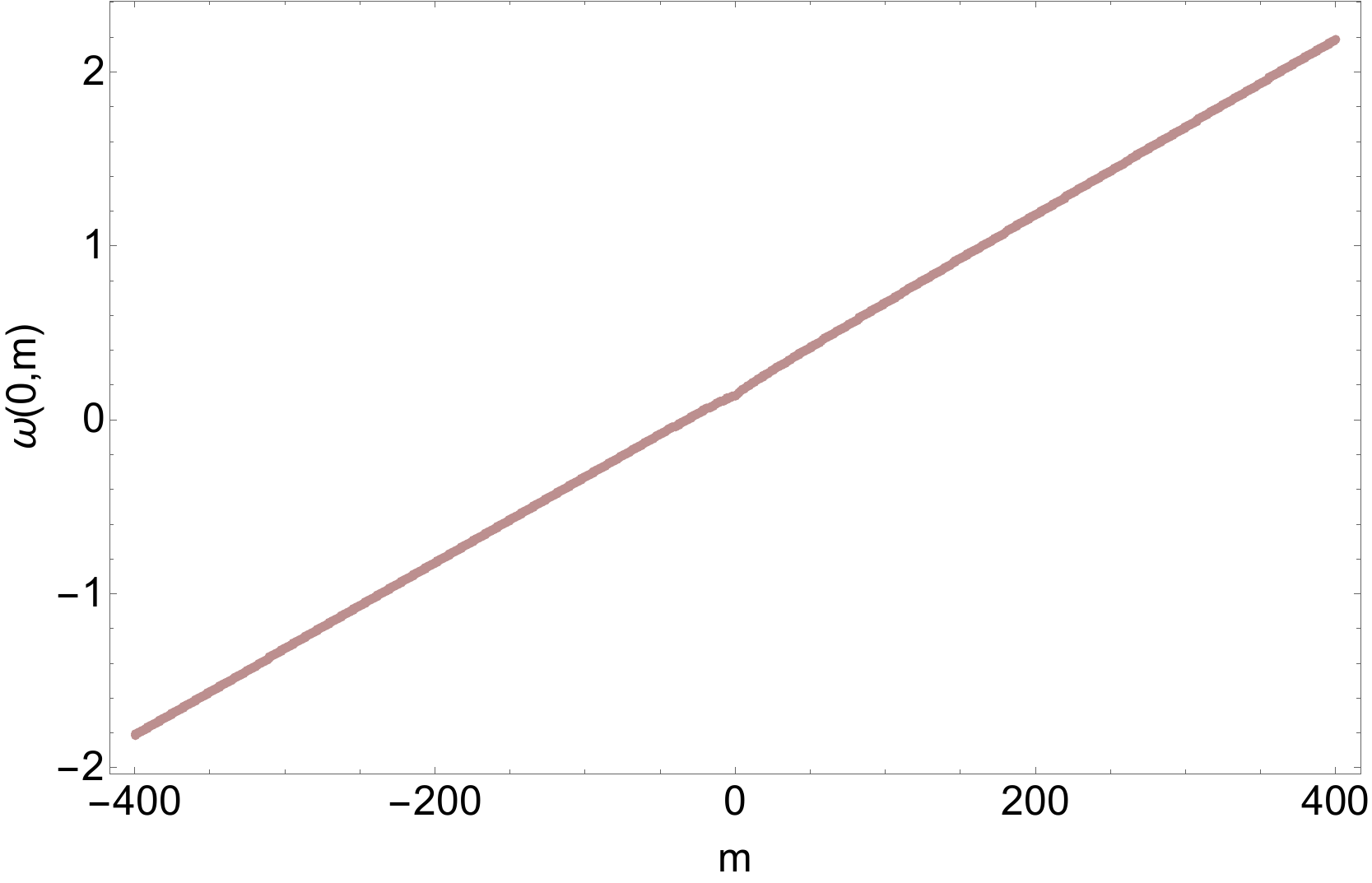}
    \end{subfigure}
    \hfill
    \begin{subfigure}{0.47\textwidth}
    \includegraphics[width=\textwidth]{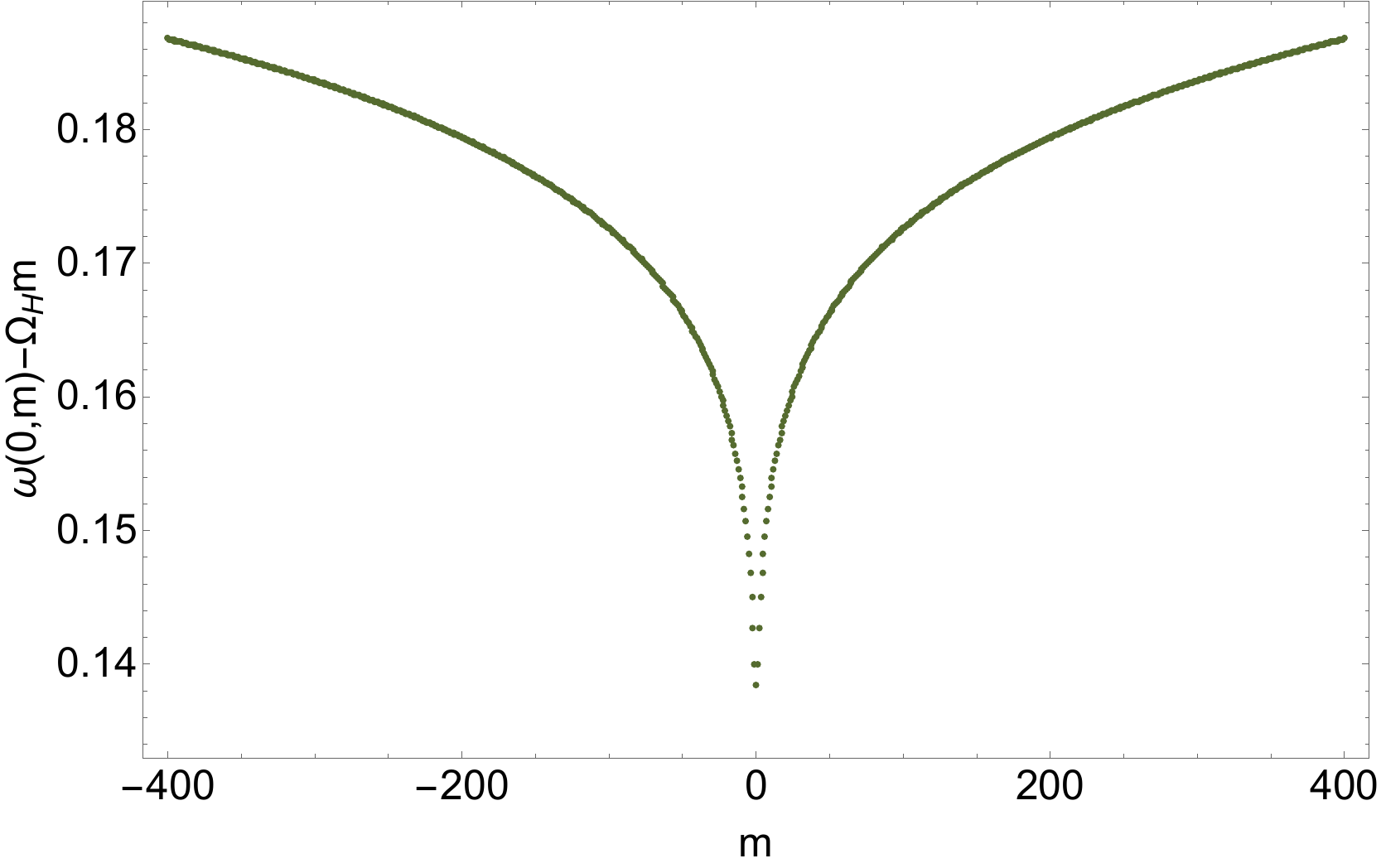}
    \end{subfigure}
    \caption{Spectrum (left) and regulated spectrum (right) for BTZ black hole with $J=0.01$. Here $n=0$, $m_{cut}=400$ and $z_0=10^{-20}$. Note that, while the unregulated spectrum appears linear, the regulated spectrum contains the qualitative features which we have seen to yield a non-trivial spectral form factor. }
    \label{J_0_01_spectrum}
\end{figure}

\begin{figure}[H]
    \centering
    \includegraphics[width=.47\textwidth]{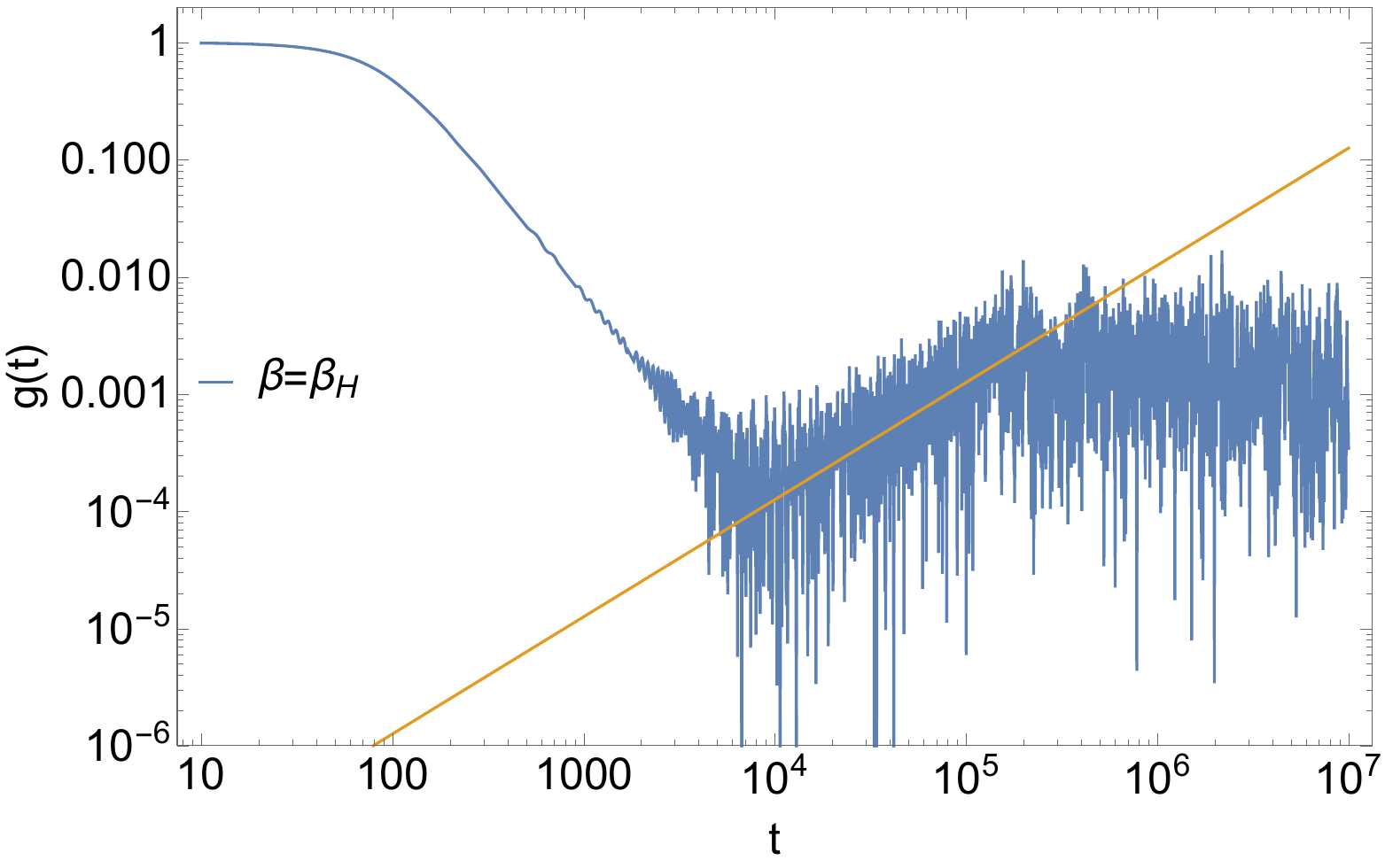}
   \caption{SFF corresponding to the modes in Figure~\ref{J_0_01_spectrum} with $\beta=\beta_H$. The yellow line has slope one.}
    \label{J_0_01_sff}
\end{figure}

\begin{figure}[H]
\begin{subfigure}{0.47\textwidth}
    \centering
    \includegraphics[width=\textwidth]{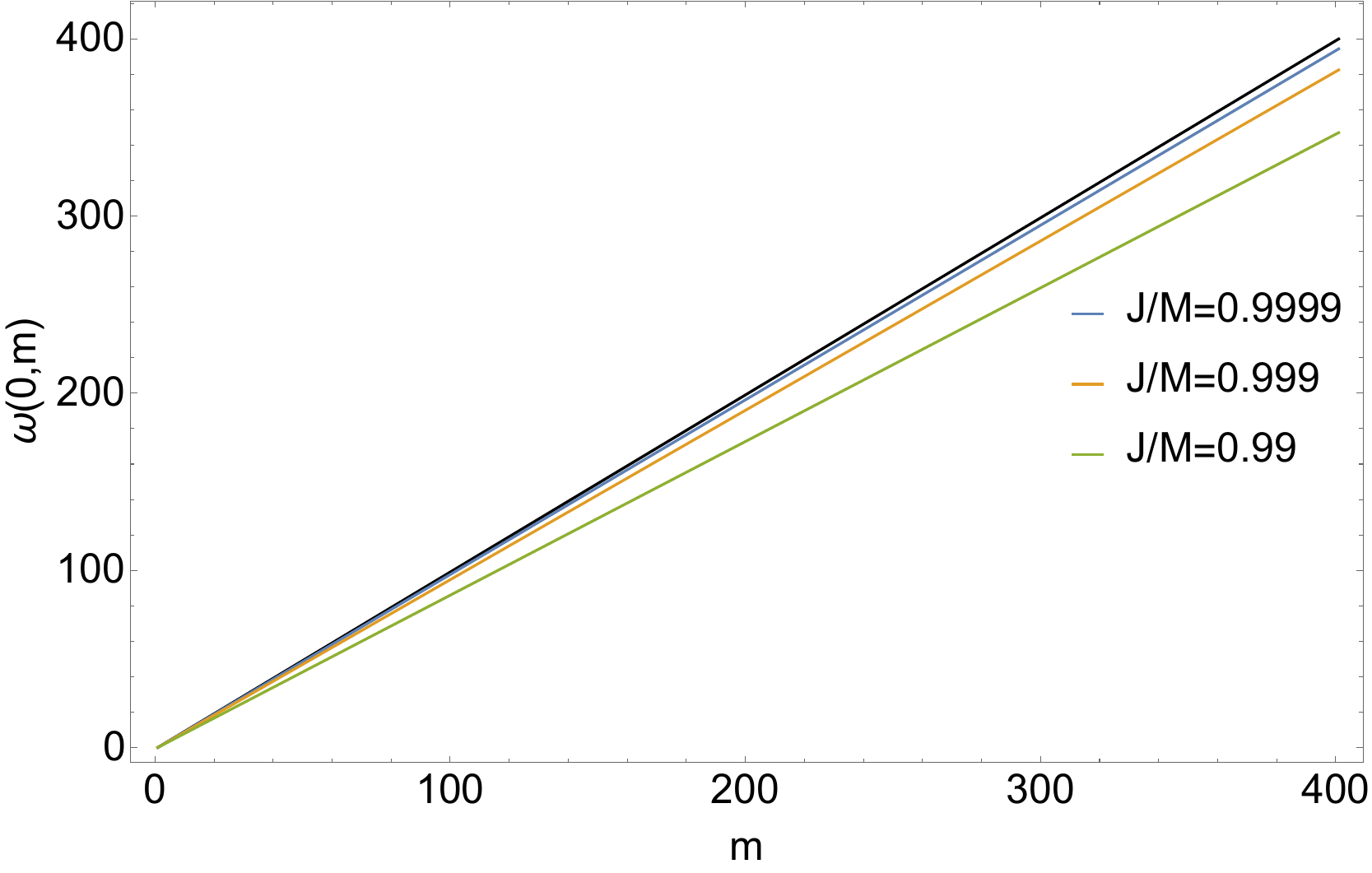}
    \end{subfigure}
    \hfill
    \begin{subfigure}{0.47\textwidth}
    \includegraphics[width=\textwidth]{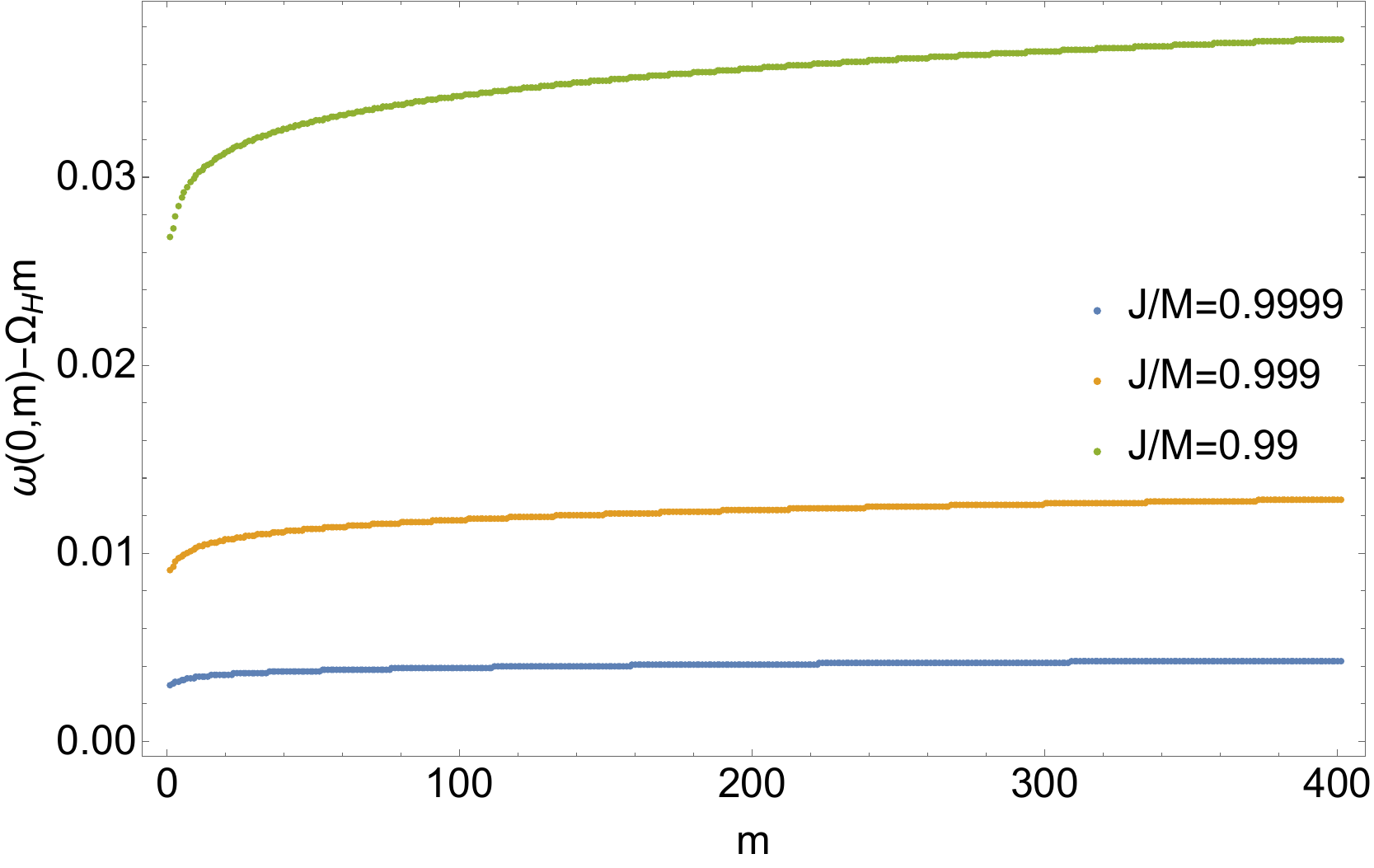}
    \end{subfigure}
    \caption{Spectrum (left) and regulated spectrum (right) for different $\frac{J}{M}$ values. Extremal limit is $\frac{J}{M}=1$ and black line in the left panel corresponds to that. Here $n=0$, $m_{cut}=400$ and $z_0=10^{-20}$.}
    \label{spectrum_comparison}
\end{figure}

We will now switch gears to the rotating black hole. For  $J \neq 0$, the defining equation \eqref{quantizn_eqn} is not symmetric under $m\rightarrow-m$. This fact is reflected in Figure~\ref{J_0_01_spectrum} (left panel), which is obtained by direct numerical solutions of the quantization conditions.  Nonetheless, the regulated $\tilde{\omega}=\omega(1,m)-\Omega_H m$ is symmetric, which is also shown in Figure~\ref{J_0_01_spectrum} (right panel).
A particularly noteworthy feature of the spectrum is: In the presence of a non-vanishing $J$, the spectrum $\omega(m)$ is not necessarily positive. To define positive modes, we can work with $\tilde{\omega}(n,m)=\omega(n,m)-\Omega_H m$. This, in turn, corresponds to considering a grand-canonical partition function and the corresponding analytically continued spectral form factor:
\begin{eqnarray}
Z\left[ \beta \right] = \sum_{m,n} e^{-\beta \omega(n,m)}  \to Z \left[ \beta, \Omega_H \right] = \sum_{m,n} e^{-\beta \left( \omega(n,m) - m \Omega_H \right)} \ , \label{gSFF}
\end{eqnarray}
where on the RHS, the partition sum consists of summing over single particle energies for each conserved angular momentum $m$, and subsequently summing over all angular momenta, weighted by a chemical potential which is fixed by the angular velocity of the event horizon. The corresponding spectral form factor is now obtained by sending $\beta \to \beta + it$, keeping $\Omega_H$ real.\footnote{In principle, it is also possible to consider analytic continuation of $\Omega_H \to \Omega_H + i s$, where $s$ is some parameter, similar to the possibility that was mentioned in \cite{Das:2022evy}. To the best of our knowledge, we are unaware of study along this direction. We will, however, not explore this possibility here. } In Figure~\ref{J_0_01_sff} we have demonstrated the generic dip-ramp-plateau behaviour of the corresponding SFF. Let us emphasize that the existence of the DRP-structure, especially the ramp with slope-one ramp, in the corresponding SFF is a robust feature. These features are similar to our earlier observations in \cite{Das:2022evy, Das:2023ulz}.
Furthermore, the left panel of Figure~\ref{spectrum_comparison} shows the behaviour of normal modes for fixed $n$ with the angular momentum $J$ of the black hole. As $J$ increases, $\omega$s become more linear with slope unity. The regulated spectrum, on the other hand, has a curvature near small angular momenta $m$; but it tends to flatten out for increasing angular momentum of the scalar field. 
%

\subsubsection{Some Analytical Regimes \& Estimates}

In this section, we will explore analytic regimes of the modes. Towards that, we will find a ``near-horizon" solution and a ``near-boundary" solution and impose a matching condition. This yields an algebraic equation which can be solved to obtain analytic expressions for the normal modes. The general solution of the equation (\ref{hyper}) can be written as:
\begin{eqnarray}
\phi(z) = z^{\frac{1}{2}(c-1)} \left[ C_1 \, {}_2F_1(a,b;c;z) + C_2 (-1)^{1-c} z^{1-c} {}_2F_1(1+a-c,1+b-c;2-c;z) \right]  \ , \label{solanal}
\end{eqnarray}
where $C_{1,2}$ are hitherto undetermined constants. Now, we wish to impose the near-horizon Dirichlet boundary condition. For this, we expand the above solution near $z=z_0$, with $z_0 \ll 1$, such that, at the leading order:
\begin{eqnarray}
\phi(z_0) = z_0^{-c/2} \left[ C_1 z_0^{c-\frac{1}{2}} +  (-1)^{1-c} C_2 z_0^{\frac{1}{2}}\right] \ ,
\end{eqnarray}
which, upon imposing $\phi(z_0) =0$, yields: $C_2 = C_1  (-1)^c z_0^{c-1}$. In deriving the above, we have explicitly used ${\rm Re}(c-1/2) = 1/2$ and therefore both terms above are on equal ground. Upon inserting the above relation in (\ref{solanal}), we end up with a solution, denoted by $\phi(z)_{\rm nh}$ with one undetermined constant. Let us now expand this solution near the boundary, as $z \to 1$. A straightforward series expansion near this point yields:
\begin{eqnarray}\label{phizhb}
&& \lim_{z\to 1}\phi(z)_{\rm nh} = \frac{\pi  C_1 \csc (\pi  (a+b-c)) \left(\frac{z_0^{c-1} \Gamma (2-c)}{\Gamma (1-a) \Gamma (1-b)} - \frac{\Gamma (c)}{\Gamma (c-a) \Gamma (c-b)}\right)}{\Gamma (a+b-c+1)} \nonumber\\
&& + (z-1)^{-a-b+c} \frac{\pi  C_1 (\cot (\pi  (a+b-c))-i) \left(z_0 \Gamma (c) \Gamma (a-c+1) \Gamma (b-c+1)-\Gamma (a) \Gamma (b) z_0^c \Gamma (2-c)\right)}{z_0 \Gamma (a) \Gamma (b) \Gamma (a-c+1) \Gamma (b-c+1) \Gamma (-a-b+c+1)} \nonumber\\
&& + \ldots 
\end{eqnarray}
Normalizability of $\phi(z)_{\rm nh}$ near $z\to 1$ imposes:
\begin{eqnarray}
z_0^{c-1} = \frac{\Gamma (1-a) \Gamma (1-b) \Gamma (c)}{\Gamma (2-c) \Gamma (c-a) \Gamma (c-b)} \ .  \label{z0rel}
\end{eqnarray}
Note that this condition is identical to (\ref{Dbchor}).

Now, using identities of hypergeometric functions, we can can rewrite $\phi(z)_{\rm nh}$ as a function of $(1-z)$. Towards this, we explicitly use:
\begin{eqnarray}
{}_2F_1(a,b,c;z) & =&  \frac{\Gamma(c) \Gamma(c-a-b)}{\Gamma(c-a) \Gamma(c-b)} {}_2F_1(a,b,a+b+1-c; 1- z) \nonumber\\
& + & \frac{\Gamma(c) \Gamma(a+b-c)}{\Gamma(a)\Gamma(b)} (1-z)^{c-a-b} {}_2F_1(c-a, c-b, 1+c-a-b; 1-z) \ . \nonumber\\
\end{eqnarray}
The resulting expression for $\phi(z)$, which we denote by $\phi(1-z)_{\rm b}$, is somewhat unwieldy, and we refrain from providing it explicitly. It is nonetheless straightforward to carry out a series expansion of $\phi(1-z)_{\rm b}$ near $z\to 1$. This yields:
\begin{eqnarray}
 \phi(1-z)_b &=& C_1 \Gamma (c) (1-z)^{-a-b+c} \Gamma (a+b-c) \nonumber\\
&& \left(\frac{1}{\Gamma (a) \Gamma (b)}-\frac{\Gamma (1-a) \Gamma (1-b)}{\Gamma (a-c+1) \Gamma (c-a) \Gamma (b-c+1) \Gamma(c-b)}\right) \nonumber\\
&& + \ldots \ ,
\end{eqnarray}
where we have used (\ref{z0rel}) above. Comparing the above with (\ref{phizhb}), we obtain the constraint:
\begin{eqnarray}
&& e^{- 2 i (a+b-c)\pi} =1 \quad \implies \quad a+b-c = n \ , \quad n\in {\mathbb Z}\ , \\
&& \implies \quad \omega = \left( \frac{r_+}{r_-}\Omega_H \right) m  \quad {\rm and} \quad n = -1 \ .
\end{eqnarray}
Note that, the formula above allows us to easily take two limits: (i) extremal: when $r_+=r_-$, we obtain $\omega = \Omega_H m$; (ii) non-rotating: $\Omega_H=0$, and $r_-=0$. It is reasonable to assume that in this limit $\Omega_H/r_-$ remains constant\footnote{In fact, it is straightforward to check that the ratio $\frac{\Omega_H}{r_-} \to \frac{1}{\ell^2 \sqrt{M}}$ in the limit $J\to 0$. } and therefore $\omega(m)$ is still a linear function.\footnote{Note that, this behaviour is expected for large values of $\omega$, measured in units of temperature. At extremality, the temperature vanishes and therefore every mode is infinitely large in this unit.} While the extremal limit yields a straightforward result, the non-rotating limit is more subtle. For this reason, we will treat this case separately.

\subsubsection{Some Analytical Regimes \& Estimates: WKB}

In this section, we will make use of the WKB-approximation to obtain the normal modes, which we have already obtained by solving the boundary conditions. The purpose here is to explore any potential analytical window as well as use numerical solutions on the WKB-equations, which are technically simpler than the boundary conditions. As before, we can obtain an explicit form of the WKB-potential, which in this case is somewhat unwieldy. Instead of presenting the detailed expression, let us discuss some instructive limits of this potential. 
\begin{figure}[H]
    \centering
    \includegraphics[width=.60\textwidth]{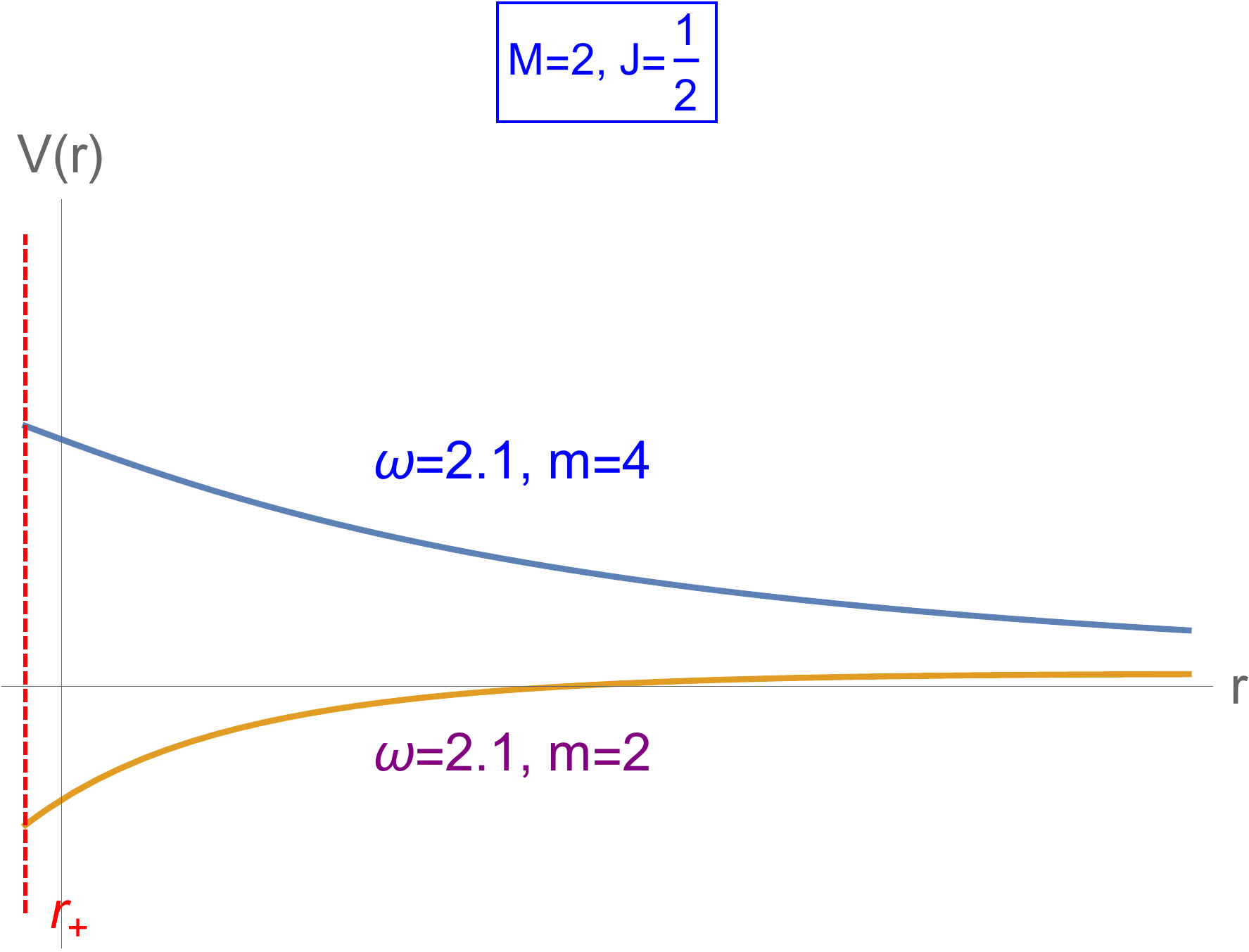}
   \caption{The Schr\"{o}dinger potential corresponding to the Klein-Gordon equation in rotating BTZ-background. There are two distinct qualitative classes of behaviours, which is shown in the blue and the dark yellow curves. The first clearly does not have any classical turning point while the latter does. We will only consider the situations with a classical turning point. The WKB-potential, in the admissible regime, has the same qualitative features that we have seen in the non-rotating BTZ-geometry as well. The location of the event horizon, $r_+$, is demonstrated by the vertical dashed line.}
    \label{wkbpot_J}
\end{figure}
First of all, it is clear from Figure~\ref{wkbpot_J} that there are two qualitatively distinct parametric regimes of the WKB potential: one which has a classical turning point and one which does not. Note, however, that this depends crucially on the allowed values of $\{\omega, m\}$ and therefore the spectrum which we are yet to determine. On physical grounds, we will consider the first case only.

Let us investigate the role of the black hole angular momentum on the spectrum, by turning on a small $\delta  = J/M \ll 1$. The WKB potential takes the form:
\begin{eqnarray}
V(r) = \frac{4 m^2 r^2-8 m^2+3 r^4-4 r^2 \omega ^2-12 r^2-4}{4 r^2\left(r^2-2\right)^2}+\frac{2 \delta  m \omega }{r^2 \left(r^2-2\right)^2} + {\cal O}(\delta^2)\ ,
\end{eqnarray}
where we have set $\ell=1$. It is straightforward to check that at $\delta=0$, the potential above reduces to (\ref{pot_noJ}). It is now somewhat tedious to perform the WKB-integral using the above potential. Nonetheless, it is possible. The detailed steps are rather messy, we therefore present the final form of the integral. Clearly, the integral $\int \sqrt{|V(r)|} dr$ consists of two terms. These yield the following two contributions:
\begin{eqnarray}
&& I = I_1 + I_2 \ , \\
&& I_1 = \frac{1}{8} \left(2 \sqrt{3} \tan ^{-1}(A_1)-\frac{\sqrt{4m^2+2} \log \left(\frac{A_2+1}{A_2-1}\right)+2\sqrt{\omega ^2+2} \log\left(\frac{A_3+1}{A_3-1}\right)+\sqrt{6} \pi}{\sqrt{2}}\right) \ , \\
&& I_2 = \frac{m \omega  \left(\frac{2 \log \left(\frac{X_1+1}{X_1-1}\right)}{\sqrt{4m^2+2}}+\frac{\log\left(\frac{X_1+1}{X_1-1}\right)}{\sqrt{\omega^2+2}}\right)}{8 \sqrt{2}}   \ , 
\end{eqnarray}
where
\begin{eqnarray}
&& A_1 = \frac{2 m^2+3 r^2-2 \omega ^2-6}{\sqrt{3} \sqrt{4 m^2 \left(2-r^2\right)-3 r^4+4 r^2 \omega ^2+12 r^2+4}} \ , \\
&& A_2 = \frac{m^2 \left(8-2 r^2\right)+2 r^2 \omega ^2+6 r^2+4}{\sqrt{2}\sqrt{4 m^2+2} \sqrt{4 m^2 \left(2-r^2\right)-3 r^4+4 r^2 \omega^2+12 r^2+4}} \ , \\
&& A_3 = \frac{r^2 \left(m^2-\omega ^2\right)-2 \left(m^2+\omega^2+4\right)}{\sqrt{2} \sqrt{\omega ^2+2} \sqrt{4 m^2\left(2-r^2\right)-3 r^4+4 r^2 \omega ^2+12 r^2+4}} \ , 
\end{eqnarray}
and
\begin{eqnarray}
&& X_1 = \frac{m^2 \left(8-2 r^2\right)+2 r^2 \omega ^2+6 r^2+4}{\sqrt{2}\sqrt{4 m^2+2} \sqrt{4 m^2 \left(2-r^2\right)-3 r^4+4 r^2 \omega^2+12 r^2+4}} \ , \\
&& X_2 = \frac{r^2 \left(m^2-\omega ^2\right)-2 \left(m^2+\omega^2+4\right)}{\sqrt{2} \sqrt{\omega ^2+2} \sqrt{4 m^2 \left(2-r^2\right)-3 r^4+4 r^2 \omega ^2+12 r^2+4}} \ . 
\end{eqnarray}

To proceed further, let us now set $M=1$ and $r = 1 + \epsilon$, with $\epsilon \ll 1$. At the leading order in both $\{\epsilon, \delta\}$, we obtain:
\begin{eqnarray}
I &=& \frac{1}{8} \left(-2 \sqrt{3} \tan ^{-1}\left(\frac{\omega^2-m^2}{\sqrt{3} \sqrt{\omega ^2+1}}\right)-\sqrt{3} \pi \right) \nonumber\\ 
&+& \frac{1}{8} \left(  \frac{\left(m \omega  \delta -2 \left(\omega ^2+1\right)\right) \log \left(\frac{\epsilon^2 \left(m^4-2 m^2 \omega^2+\omega ^4+3 \omega ^2+3\right)}{4 \left(\omega^2+1\right)^2}\right)}{\sqrt{\omega ^2+1}} \right) \nonumber\\
&+& \frac{1}{8} \left(  \frac{\log \left(\frac{\sqrt{4 m^2+1} \sqrt{\omega ^2+1}+m^2+\omega^2+2}{-\sqrt{4 m^2+1} \sqrt{\omega ^2+1}+m^2+\omega ^2+2}\right) \left(-4 m^2+2 m \omega  \delta -1\right)}{\sqrt{4 m^2+1}} \right) \ . 
\end{eqnarray}
Now an analytic expression for the normal modes can be obtained by solving the WKB-equation in the limit $m \gg \omega$ and $m \delta \ll 1$, which yields:
\begin{eqnarray}
\omega = \sqrt{\frac{4 \pi ^2 a^2}{\log \left(\frac{2}{m^2 \epsilon}\right)}-1}+\frac{\delta  m}{2} + \ldots \ , \quad a = \frac{3}{4} + 2 n \ . \label{omegaexact_J}
\end{eqnarray}
It is straightforward to check that when $\delta=0$ and $m^2\epsilon \ll 1$ the formula in (\ref{omegaexact_J}) matches with (\ref{omegaexact}). Note that, (\ref{omegaexact_J}) is already suggestive about the role of a non-vanishing $J$ in the spectrum: First, the leading order contribution already linearizes $\omega(m)$. It is therefore possible that at the maximal $J$, the linear modes will dominate. However, this needs to be checked separately, which we will do in the next section. Secondly, it is manifest that $(\omega - c J m)$ behaves as the modes in (\ref{omegaexact}), where $c$ is an order one constant. Thus, an appropriately defined SFF in terms of $(\omega - J m)$ is expected to display the robust DRP-structure that we have previously seen. Note that, this consideration directly leads to the SFF obtained from the grand-canonical partition function, as already mentioned in (\ref{gSFF}).

\subsection{Extremal BTZ: $J=M$ case} 

Let us consider the special case of the extremal limit, which is obtained by setting $J = M \ell$ in (\ref{metric}). Evidently, this sets $r_+ = r_-$. Let us write the corresponding metric as follows:
\begin{eqnarray}\label{metextremal}
    ds^2=-f(\rho)dt^2+\frac{d\rho^2}{f(\rho)} + \rho^2\left(dy-\frac{r_H^2}{\rho^2}dt\right)^2 \ ,  \quad \text{with}\hspace{0.5mm} \quad  f(\rho)=\frac{(\rho^2-r_H^2)^2}{\rho^2} \ , \label{metextremal_text}
\end{eqnarray}
where $r_H = r_+ = r_-$, $\rho \in [r_H, \infty]$ is the radial coordinate and $y \sim y + 2\pi R_y$ is the compact direction.

Before moving further, let us offer a few comments on the extremal geometry. The extremal BTZ geometry can be written in the following form:
\begin{eqnarray}
&& ds^2 = r_H^2 du_-^2 + dx^2 +  e^{2x} du_+ du_-  \ , \label{extnh} \\
&& u_\pm = y \pm t \ , \quad \rho^2 - r_H^2 = e^{2x} \ , 
\end{eqnarray}
The coordinates $u_\pm$ are periodic: $u_\pm \sim u_\pm + 2\pi$. The boundary is located at $x\to \infty$, where $u_\pm$ become null directions; the horizon is located at $x\to - \infty$. It is instructive to compare the near-horizon geometry of a non-extremal black hole, {\it i.e.}~a Rindler geometry, with the metric in (\ref{extnh}). While the non-trivial warp factor $e^{2x}$ fast approaches zero near the horizon (as $x\to -\infty$), the extremal geometry in (\ref{extnh}) retains a constant $du_-^2$ deformation supported by $r_H^2$ in the first term of the metric. This is qualitatively different from a Rindler$\times S^1$ geometry and explicitly can be written as an ${\rm AdS}_2\times S^1$-background. This will be crucial in a qualitatively different physics at the extremal point.

For convenience, we will redefine a radial coordinate: $r^2 = \rho^2 - r_H^2$ and rewrite  \eqref{metextremal} in the following form:
\begin{equation}\label{met1}
    ds^2=-f(r)dt^2+\frac{r^2}{f(r)(r^2+r_H^2)}dr^2+(r^2+r_H^2)\left(dy^2-\frac{r_H^2}{r^2+r_H^2}dt \right)^2, \hspace{0.3cm} \text{with}\hspace{0.2cm}  f(r)=\frac{r^4}{r^2+r_H^2}
\end{equation}
where $r=0$ is horizon and $r\rightarrow \infty$ is the conformal boundary. With the decomposition of scalar field $\Phi(t,,r,y)\sim e^{-i \omega t} e^{i m y} \phi(r)$, KG equation takes the following form,
\begin{equation}\label{extrad}
    \phi(r) \left(m^2 \left(r_H^2-r^2\right)-2 m r_H^2 \omega -\mu ^2 r^4+\omega ^2 \left(r^2+r_H^2\right)\right)+r^5 \left(r \phi''(r)+3 \phi'(r)\right)=0.
\end{equation}
Let's define two new variables $p$ and $q$ as $p=m+\omega$ and $q=m-\omega$. In terms of $z=\frac{i r_H q}{r^2}$, \eqref{extrad} is a confluent hypergeometric equation,
\begin{equation}
    \phi''(z)+\left(-\frac{\mu^2}{4z^2}+\frac{i p}{4 r_H z} -\frac{1}{4}  \right)\phi(z)=0,
\end{equation}
with solution is in terms of  Whittaker functions,
\begin{equation}\label{whit1}
    \phi(z)=C_1 M_{\kappa,\nu}(z)+C_2 W_{\kappa,\nu}(z), \hspace{0.9cm}\text{with}\hspace{0.2cm}  \kappa=\frac{i p}{4 r_H}, \, \nu=\frac{\sqrt{1+\mu^2}}{2}
\end{equation}
 Near boundary ($z\rightarrow 0$) expansion of \eqref{whit1} is,
\begin{equation}
    \phi_{\rm bdry}(z)\sim (C_1+A \, C_2) z^{\frac{1}{2}(1+\sqrt{1+\mu^2})}+ C_2 \, B z^{\frac{1}{2}(1-\sqrt{1+\mu^2})},
\end{equation}
where first term is normalizable and second one is non-normalizable. So normalizability of the scalar field near boundary implies $C_2=0$, i.e. $\phi(z) \sim  M_{\kappa,\nu}(z) $ whose near horizon ($z\rightarrow \infty$) behaviour is the following,
\begin{equation}
    \phi_{\text{hor}}(z) \sim \Gamma(\Delta) \left(\frac{e^{\frac{z}{2}}z^{-\frac{i p}{4r_H}}}{\Gamma(\frac{\Delta}{2}-\frac{ip}{4r_H})} +(-1)^{\frac{ip}{4r_H}-\frac{\Delta}{2}} \frac{e^{-\frac{z}{2}}z^{\frac{ip}{4r_H}}}{\Gamma(\frac{\Delta}{2}+\frac{ip}{4r_H})} \right).
\end{equation}
Dirichlet boundary condition near horizon that $ \phi_{\text{hor}}(z=z_0)=0$ implies,
\begin{equation*}
    \begin{split}
        \frac{\Gamma(\frac{\Delta}{2}+\frac{ip}{4r_H})}{\Gamma(\frac{\Delta}{2}-\frac{ip}{4r_H})} &=  -(-1)^{\frac{ip}{4r_H}-\frac{\Delta}{2}} e^{-z_0} z_0^{\frac{ip}{2r_H}}\\
        &= -e^{\frac{p\pi}{4 r_H}+\frac{i\pi \Delta}{2}} e^{-z_0} \left(\frac{ir_H q}{\epsilon^2} \right)^{\frac{ip}{2r_H}} \\
        &= -e^{\frac{p\pi}{4 r_H}+\frac{i\pi \Delta}{2}} e^{-z_0} \left(\frac{r_H q}{\epsilon^2} \right)^{\frac{ip}{2r_H}}  e^{-\frac{\pi p}{4 r_H}}\\
        &= -e^{\frac{i\pi \Delta}{2}} e^{-\frac{i r_H q}{\epsilon^2}} \left(\frac{r_H q}{\epsilon^2} \right)^{\frac{ip}{2r_H}}
    \end{split}
\end{equation*}
\begin{equation}
    \Rightarrow   \frac{\Gamma(\frac{\Delta}{2}+\frac{ip}{4r_H})}{\Gamma(\frac{\Delta}{2}-\frac{ip}{4r_H})}   e^{\frac{-i\pi \Delta}{2}} e^{\frac{i r_H q}{\epsilon^2}} \left(\frac{r_H q}{\epsilon^2} \right)^{-\frac{ip}{2r_H}}  =-1
\end{equation}
Which leads to the following quantization condition,
\begin{equation}\label{quant2}
    2 \text{Arg} \left[  \Gamma \left(\frac{\Delta}{2}+\frac{ip}{4r_H} \right)  \right] + \frac{r_H q}{\epsilon^2}-\frac{\pi \Delta}{2} + \frac{p}{2r_H} \log \left(\frac{\epsilon^2}{r_H q}\right)= (2n+1) \pi,  \hspace{1cm} n\in \mathbf{Z}.
\end{equation}
Where $\epsilon^2=\frac{i r_H q}{z_0}$. When $\epsilon \rightarrow 0$ i.e. the position of stretched horizon is very close to the event horizon, we can approximately write  \eqref{quant2} as,
\begin{eqnarray}
    & \frac{r_H q}{\epsilon^2} \sim (2n+1) \pi \nonumber  \\
     \Rightarrow &  \omega(n, m) \sim m- (2n+1) \pi \epsilon^2
\end{eqnarray}
So in the extremal limit, normal modes $\omega$ are linear in both $m$ and $n$, especially the slope of $\omega(m)$ vs. $m$ for fixed $n$ is $1$.

\subsection{Extremal Limit: WKB Approximation}

\begin{figure}[H]
    \centering
    \includegraphics[width=.60\textwidth]{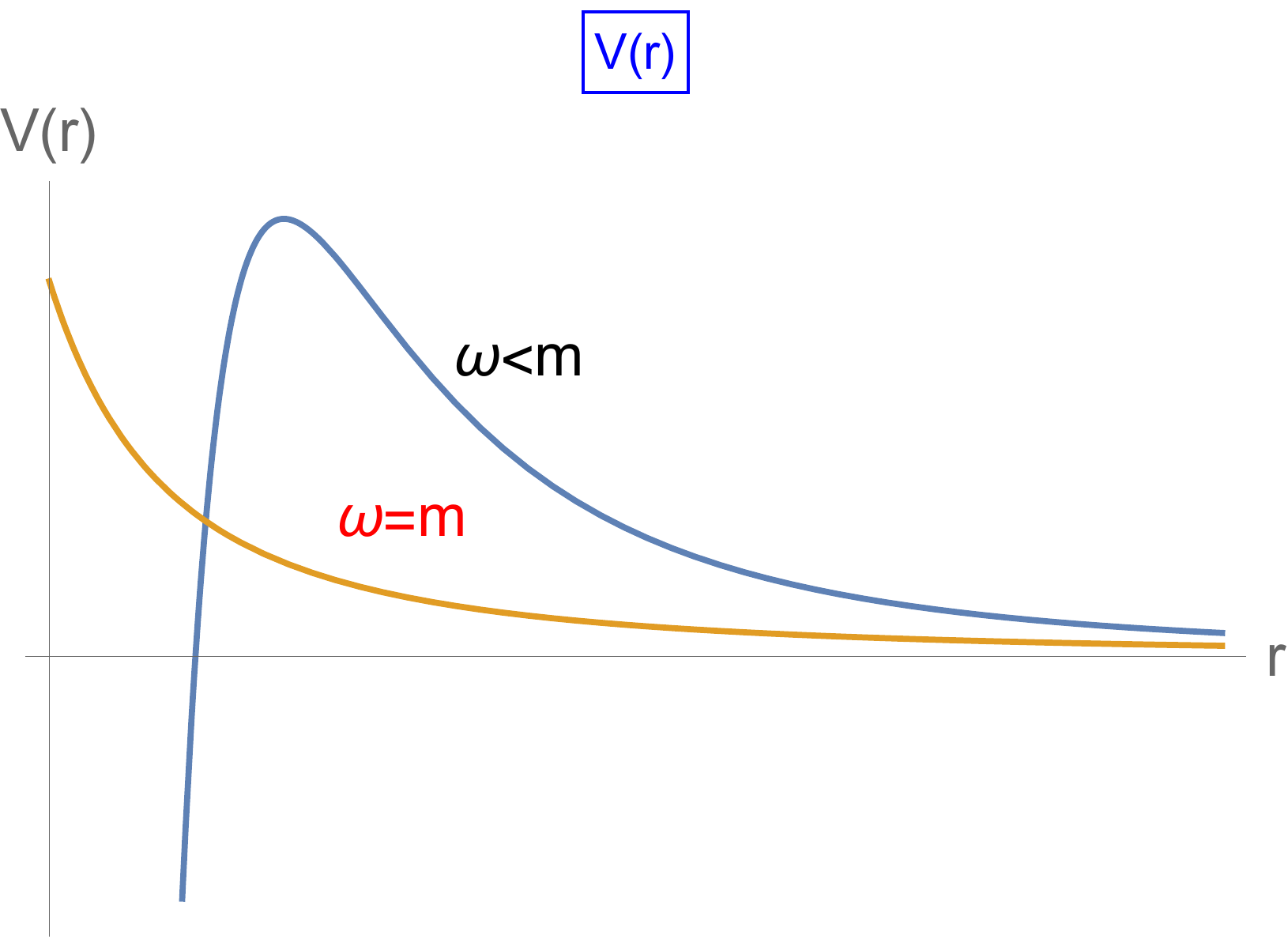}
   \caption{The Schr\"{o}dinger potential corresponding to the Klein-Gordon equation in extremal BTZ-background. There are two distinct qualitative classes of behaviour: $\omega \not =m$, which is shown in the blue curve (we have chosen $\omega =1.6$, $m=2$ for a representative value). The other class corresponds to setting $\omega =m$, which is shown by the dark yellow curve. The latter clearly does not have any classical turning point. We will discard this possibility explicitly. The WKB-potential, in the admissible regime, has the same qualitative features that we have seen in the non-rotating BTZ-geometry as well.}
    \label{wkbpot_extremal}
\end{figure}

In order to carry out the WKB approximation, we will rewrite the Klein-Gordon equation \eqref{extrad} in terms of a Schr\"{o}dinger equation of the following form:
\begin{eqnarray}
&& \frac{d^2 \Psi(r)}{dr^2} - V(r) \Psi(r) = 0 \ , \\
&& V(r) = \frac{4 m^2 \left(r^2 - r_H^2\right) + 8 m r_H^2 \omega +3 r^4 - 4 r^2 \omega ^2 - 4 r_H^2 \omega ^2}{4 r^6} \ . 
\end{eqnarray}
The classical turning points of the above potential is given by: 
\begin{eqnarray}
V(r_c)=0 \quad \implies \quad r_c^2=\frac{2}{3}\left( \omega^2 - m^2 + (\omega-m)\sqrt{3+(\omega+m)^2}\right)  \ .
\end{eqnarray}
It is clear that $\omega=m$ is a special parametric locus, which does not have a turning point. We have pictorially demonstrated the qualitative features of the potential in figure~\ref{wkbpot_extremal}. 

Given the potential, we can evaluate the integral on the LHS of (\ref{wkbrule}). This yields:
\begin{eqnarray}
&& {\rm LHS} = \frac{1}{8} \left( a + 2\sqrt{3} \arctan(b) + (m+\omega) \log(c) \right) \ , \\
&& a = \frac{2 \sqrt{-4 m^2 \left(\epsilon ^2-1\right)-8 m \omega -3 \epsilon ^4+4 \omega ^2 \left(\epsilon ^2+1\right)}}{\epsilon ^2}-\sqrt{3} \pi \ , \\
&& b = \frac{2 m^2-2 \omega ^2+3 \epsilon ^2}{\sqrt{3} \sqrt{-4 m^2 \left(\epsilon ^2-1\right)-8 m \omega -3 \epsilon ^4+4 \omega ^2 \left(\epsilon ^2+1\right)}} \ , \\
&& c = \frac{\frac{m \left(\epsilon ^2-2\right)+\omega  \left(\epsilon ^2+2\right)}{\sqrt{-4 m^2 \left(\epsilon ^2-1\right)-8 m \omega -3 \epsilon ^4+4 \omega ^2
   \left(\epsilon ^2+1\right)}}+1}{\frac{m \left(\epsilon ^2-2\right)+\omega  \left(\epsilon ^2+2\right)}{\sqrt{-4 m^2 \left(\epsilon ^2-1\right)-8 m \omega -3 \epsilon
   ^4+4 \omega ^2 \left(\epsilon ^2+1\right)}}-1} \ ,
\end{eqnarray}
where $\epsilon$ denotes the location of the cut-off surface. In the limit $\epsilon \to 0$, we an simply keep the $\epsilon^{-2}$-term coming from $a$ and ignore the rest. This yields:
\begin{eqnarray}
{\rm Abs}(\omega - m) = 2\pi \epsilon^2 \left( \frac{3}{4} + 2 n \right) \quad \implies \quad \omega = m \pm 2\pi \epsilon^2 \left( \frac{3}{4} + 2 n \right) \ , n \in {\mathbb Z}. \ . \label{omega_extremal_exact}
\end{eqnarray}
Note that, in the strict $\epsilon=0$ case, the above analysis breaks down since there are no classical turning points in this limit.

\begin{figure}[H]
     \centering
     \begin{subfigure}[b]{0.4\textwidth}
         \centering
         \includegraphics[width=\textwidth]{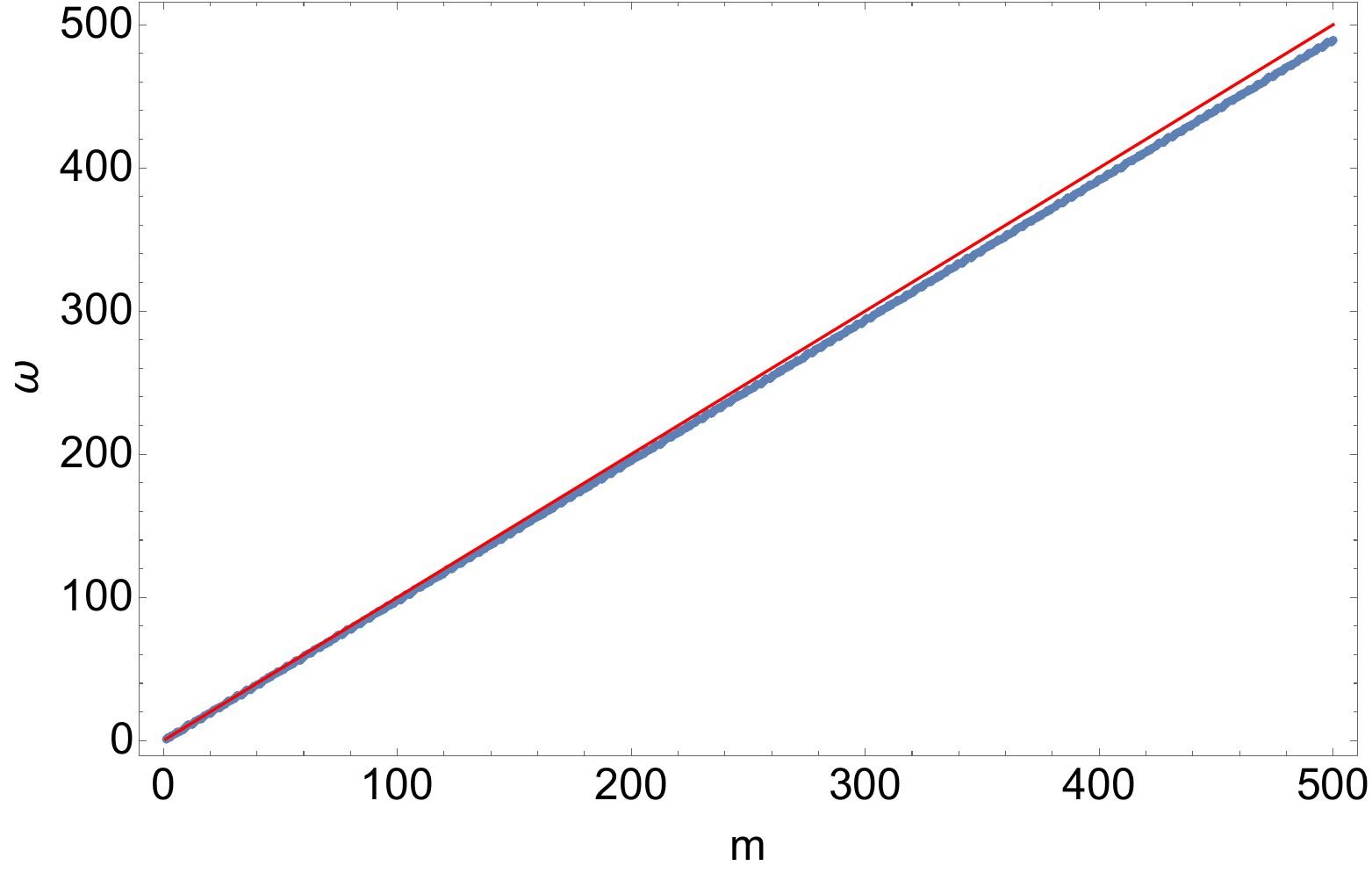}
         \caption{$\epsilon=0.1$}
         \label{fig_1_1}
     \end{subfigure}
     \hfill
      \begin{subfigure}[b]{0.4\textwidth}
         \centering
         \includegraphics[width=\textwidth]{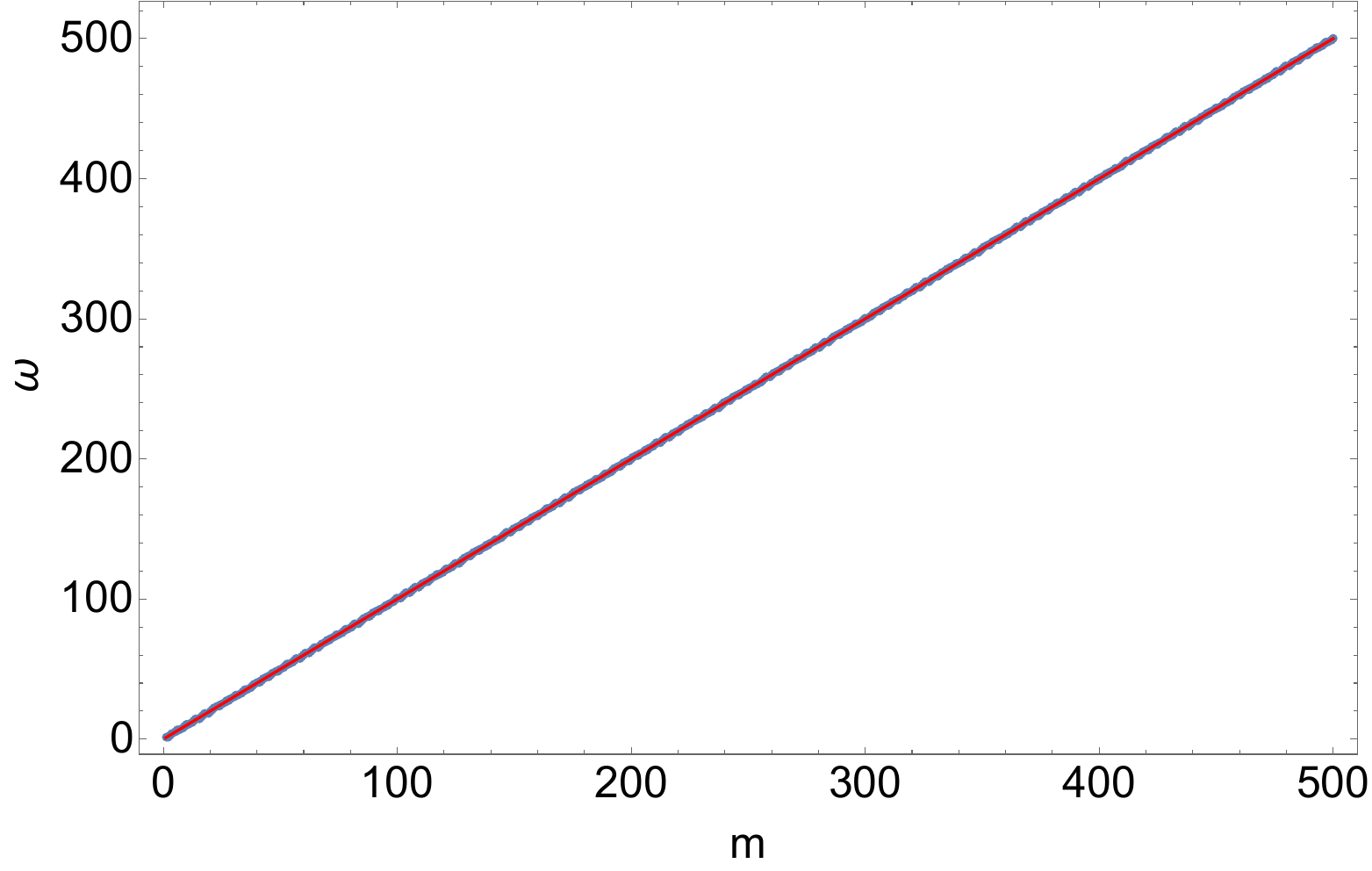}
         \caption{$\epsilon=0.001$}
         \label{fig_1_001}
     \end{subfigure}
     \hfill
     \begin{subfigure}[b]{0.4\textwidth}
         \centering
         \includegraphics[width=\textwidth]{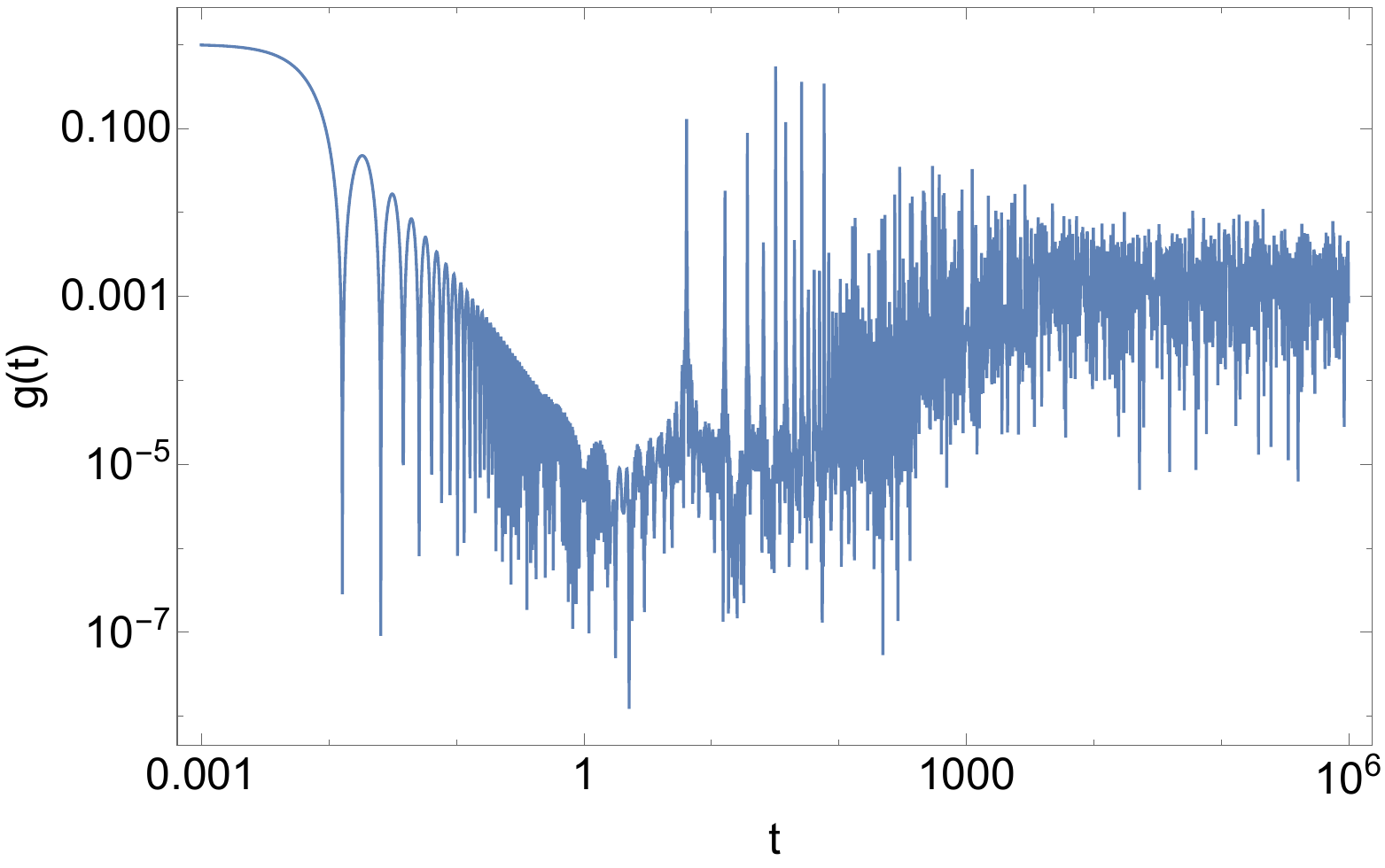}
         \caption{$\epsilon=0.1$, $\beta=0$.}
         \label{fig_1_10}
     \end{subfigure}
     \hfill
  \begin{subfigure}[b]{0.4\textwidth}
         \centering
         \includegraphics[width=\textwidth]{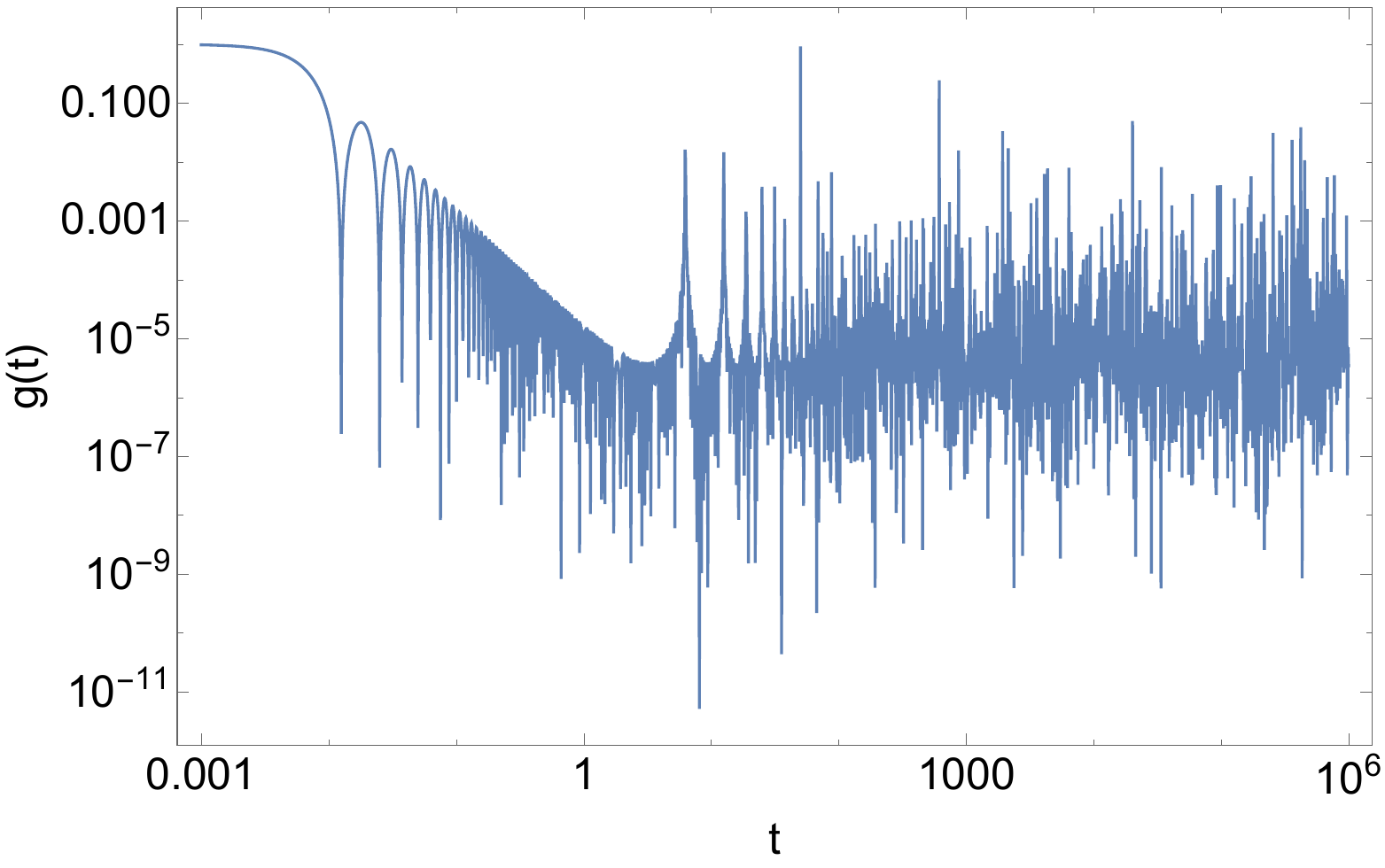}
         \caption{$\epsilon=0.001$, $\beta=0$.}
         \label{fig_1_0010}
     \end{subfigure}
     \hfill
     \begin{subfigure}[b]{0.4\textwidth}
         \centering
         \includegraphics[width=\textwidth]{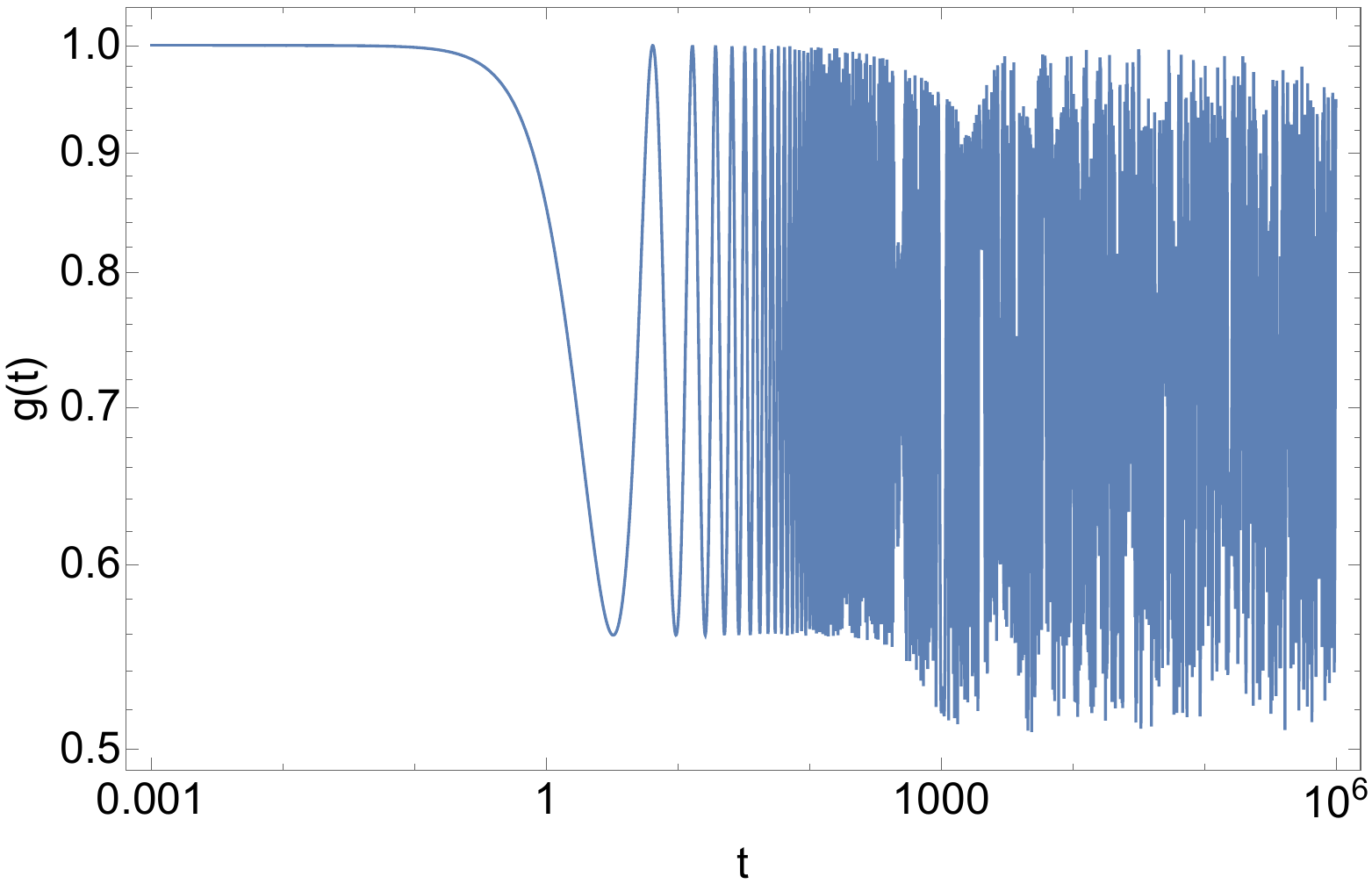}
         \caption{$\epsilon=0.1$, $\beta=2$.}
         \label{fig_1_12}
     \end{subfigure}
     \hfill
     \begin{subfigure}[b]{0.4\textwidth}
         \centering
         \includegraphics[width=\textwidth]{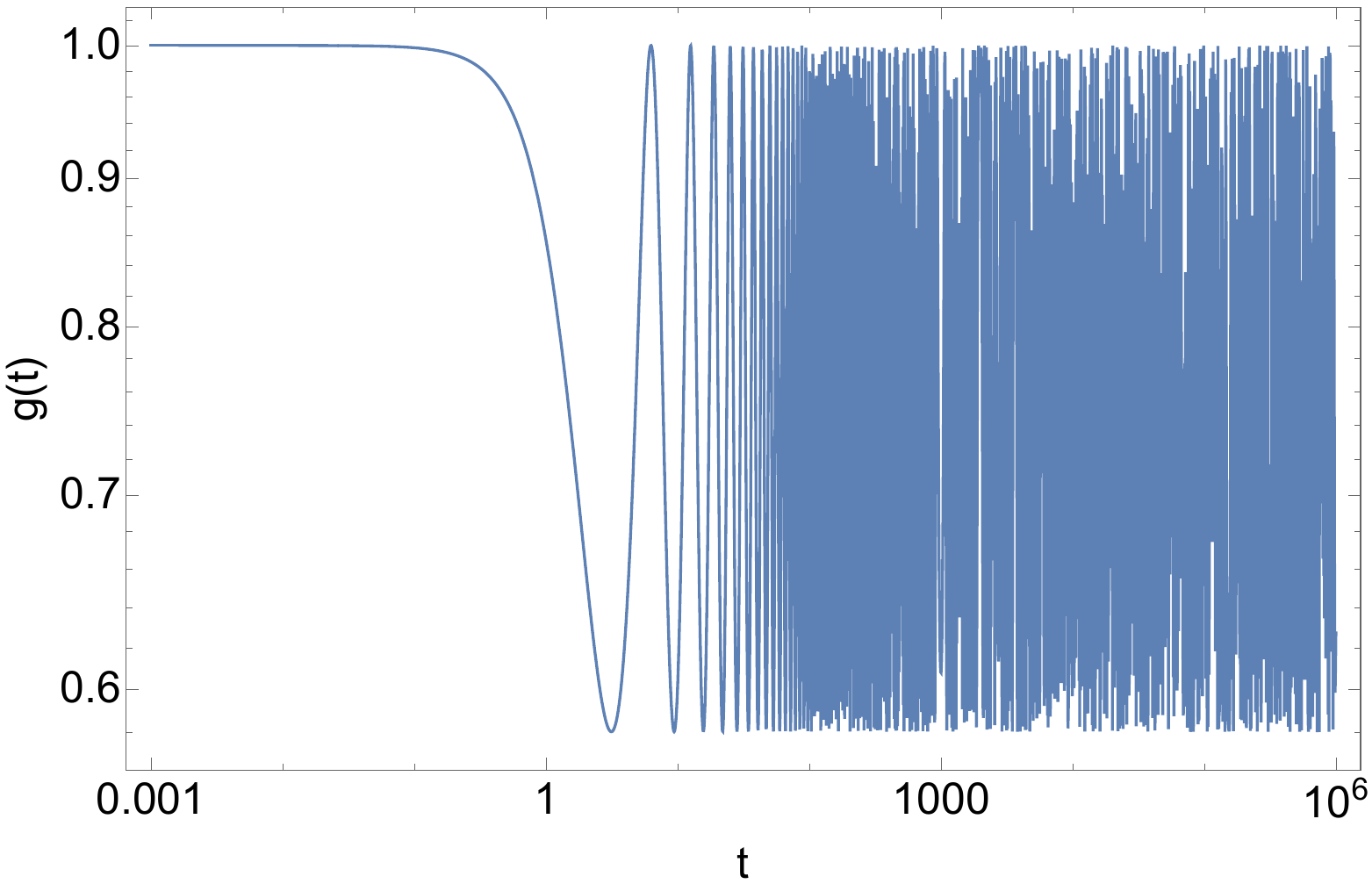}
         \caption{$\epsilon=0.001$, $\beta=2$.}
         \label{fig_1_0012}
     \end{subfigure}
        \caption{A pictorial representation of numerical solutions of the WKB-equations. We have shown the behaviour of the modes, as well as the corresponding SFF, for representative choices of $\epsilon$ and $\beta$. }
        \label{extremal_numerical}
\end{figure}

Let us demonstrate the behaviour of numerical solutions of the WKB-equations, which is summarized in Figure~\ref{extremal_numerical}. In Figures~\ref{fig_1_1} and \ref{fig_1_001} we have shown the behaviour of the normal modes $\omega(m)$, for two choices of $\epsilon=10^{-1}$ and $\epsilon=10^{-3}$, respectively. It is evident that the function $\omega(m)$ is tantalizingly close to a linear one with unit, which becomes stronger as $\epsilon$ is reduced. This observation is in qualitative agreement with the analytical formula in (\ref{omega_extremal_exact}). In Figures~\ref{fig_1_10} and \ref{fig_1_0010} the corresponding SFFs are shown with $\beta=0$. The discernible ramp-structure has disappeared here. To further emphasize this point, we have shown SFFs for $\beta=2$ in Figures~\ref{fig_1_12} and \ref{fig_1_0012},\footnote{Note that, at exact extremality, the black hole temperature vanishes and therefore a natural choice of $\beta = T^{-1} \to \infty$. Choosing $\beta=0$ is maximally far away from this parametric regime. It is highly instructive that even in this limit we do not observe any ramp structure. On the other hand, an order one value of $\beta$ makes the SFF resemble a harmonic oscillator, with a very small classical Poincar\'{e} recurrence time. These are strong hints that we loose the chaotic features at exact extremality.} which looks very similar to a harmonic oscillator SFF. We take these observations to conclude that at exact extremality, the SFF displays integrable behaviour rather than a chaotic one. 
%

\section{Comments on Back-reaction}

In this section, we will comment on the validity of the probe limit. Our analyses will be based on a classical consideration only and the physics point that we want to highlight here is that modes with arbitrarily large $m$ quantum number will back-react significantly on the classical geometry and change the metric altogether. Therefore, it is sensible to cut-off the partition sum at a finite value of $m$. We will only discuss some salient features here and refer the interested Reader to appendix \ref{probe_appendix} for more technical details.

The classical stress-tensor of the scalar field reads:
\begin{eqnarray}
T_{\alpha\beta} = \partial_\alpha \Phi \partial_\beta \Phi -  \frac{1}{2} g_{\alpha \beta} \left( \partial \Phi \right)^2 \ . 
\end{eqnarray}
it is straightforward to evaluate the corresponding energy density $\rho = T_{\alpha \beta} u^\alpha u^\beta$, where $u^\alpha$ is a time-like unit vector. Given the BTZ-geometry, the corresponding energy density behaves as (see equations (\ref{backbdry}) and (\ref{energy1})):
\begin{eqnarray}
\rho_{\rm bdry} \sim {\cal O}(r^{-6}) + \ldots \ , \quad {\rm as} \quad r\to \infty \\
\rho_{\rm hor} \sim \frac{1}{r^2 -1} + {\rm oscillatory} \ , \quad {\rm as} \quad r\to 1 \ .
\end{eqnarray}
It is clear from the expressions above that near the boundary it is always safe to ignore the back-reaction, however, near the horizon this is not the case. Ignoring the rapidly oscillatory terms above, it is straightforward to estimate the diverging back-reaction as one approaches the horizon. From (\ref{back1}) and (\ref{back2}), the back-reaction is controlled by the parameter $m/\delta$, where $\delta = r_0 - r_H$, and we have set $r_H=1$ here. Thus, our calculations are valid till $m_{\rm cut} \sim \delta^{-1}$ and this justifies the computation of a partition function with an explicit cut-off. This behaviour can simply be fixed from dimensional analysis, if we assume that there is no other scale when the brickwall is Planck distance away from the classical event horizon. It is, however, interesting that here this relation holds for arbitrary $\delta$.

Before concluding, let us offer some further comments. Note that, the above conclusion is based solely on the classical back-reaction. It is known that a semi-classical (quantum) back-reaction of a scalar field can produce a qualitatively different back-reaction, see {\it e.g.}~\cite{Casals:2016ioo, Casals:2016odj, Casals:2019jfo, Emparan:2020znc}. Therefore it merits a separate and thorough analyses to make a stronger claim regarding cutting off the large $m$ spectrum. We leave this for a future work.

\section{Discussions}

In this article, we have explored the brickwall-like model for a rotating BTZ geometry, especially in the context of a probe scalar single-particle spectral form factor. We have explicitly demonstrated that, for a generic value of the black hole mass and angular momentum, the dip-ramp-plateau structure, with a ramp of slope unity, is a stable feature of this sector. However, the crucial observation is that the non-trivial SFF dynamics is visible using the grand-canonical partition function, as opposed to the canonical partition function. This grand-canonical partition function is obtained by summing over each angular momentum sector of the probe field, in which the angular momenta of the rotating BTZ geometry plays the role of a chemical potential. Clearly, in the vanishing limit of this rotation, the grand-canonical partition function reduces to the canonical partition which we have explored in \cite{Das:2022evy}.

Furthermore, we have also demonstrated that close to extremality, {\it i.e.}~as $J \to M \ell$ this structure appears to be a robust feature still. We have used two main technical methods: A direct and brute force numerical solution of the Dirichlet boundary conditions and the WKB-approximation. We have explicitly shown that the WKB-approximation is technically easier for finding the normal modes, and it can also provide us with analytical answers in a suitable regime of the parameter space. The latter is useful for several reasons: the central theme is that it allows for an analytical understanding of the origin of the level-repulsion dynamics of the modes, which is crucial to yield the DRP-structure, especially the ramp, of the SFF.

In contrast to the above, at exact extremality, {\it i.e.}~at $J = M \ell$, the DRP structure appears lost from the SFF. At the same time, the corresponding level-spacing distribution also resembles that of a harmonic oscillator.  Thus the system becomes integrable. At extremality we have $T=0$ identically, where $T$ is the temperature of the dual CFT. Naively, it is therefore expected that any chaotic signature of the underlying dynamics will become invisible at this point. On the other hand, the dual CFT has two distinct temperature for a rotating BTZ geometry: The right-moving temperature $T_+ \sim (r_+ - r_-)$ and the left-moving temperature $T_- \sim(r_+ + r_-)$. At extremality, the right temperature vanishes but the left one remains non-vanishing. It is, therefore, an interesting question to explore how the SFF can see the non-vanishing left temperature. This question is especially interesting since it is known that an ``early time chaos", around the scrambling time-scale, can be observed with the left temperature on a suitably chosen four-point out-of-time-order correlator, see {\it e.g.}~\cite{Banerjee:2018twd, Banerjee:2018kwy, Banerjee:2019vff}. It may be possible to formulate an OTOC-calculation in our framework using a boundary CFT approach, similar to the ones in \cite{Das:2019tga, Das:2021qsd}. Alternatively, a bulk observer localized to move along the stretched horizon may also describe a suitable CFT evolution, by identifying the bulk integral curve generated by the bulk observer with a Hamiltonian evolution on the boundary CFT\cite{Das:2021qsd}. We hope to address this issue in future.

Our observations, notwithstanding, raise several further questions. First, it is not understood how one should think about the cut-off surface located at a bulk point in the geometry, from the perspective of the dual CFT. In other words, our construction appears agnostic about the specific UV-complete description and it is not clear to us how this aspect affects the robust and universal observation that we have made. One way to address this question is perhaps to consider a framework similar to the $T{\bar T}$-bar deformation where such a bulk cut-off surface emerges, following {\it e.g.}~\cite{Guica:2019nzm}, and re-interpret the Dirichlet hypersurface as an IR cut-off.\footnote{We thank Monica Guica for a conversation related to this.} Although from a gravitational point of view, this appears possible, it remains to be seen what this means in terms of the dual CFT. It would be rather interesting if the corresponding operation to the CFT can be identified with fuzzball-like states. We hope to address this question in near future.

A related question is to understand how a Dirichlet boundary condition can lead to, at least, a thermal-like CFT correlator. The universal presence of the DRP-structure hints that a corresponding thermal behaviour is expected, viewed from the dual CFT. It is thus interesting to re-visit the standard AdS/CFT correlation function calculations, in the presence of the Dirichlet surface in front of the horizon. Work along this direction is underway\cite{Das:TBA} and we hope to report on it soon. A related class of similar questions can be explored for de-Sitter space as well, which is also about to appear soon\cite{Pradipta}.

Note that, by construction, our model is at best an effective one and is not expected to provide answers to all fine-grained questions on the ramp emerges. Another way of probing the UV-complete description of the DRP-structure is to investigate a full-fledged fuzzball geometry. By now, there is a huge literature on such geometries, which are constructed from explicit supergravity solutions. While these geometries are rather rich in structure, recent advances in \cite{Bena:2019azk} obtain explicit correlation functions in such geometries and therefore greatly facilitates an SFF-computation. Especially since, at a technical level, WKB-approximation is extremely useful in such questions in the fuzzball-geometries. Building on our earlier as well as current work, we are further exploring this question in a specific class of fuzzball geometries\cite{wip}.

Note that, in the series of studies that we initiated in \cite{Das:2022evy}, followed by \cite{Das:2023ulz, Das:2023yfj}, we have considered the SFF obtained from the single-particle partition function. In other words, we have explicitly set the density of states to be unity. However, non-trivial physics is expected to result from a non-trivial density of states. For example, if we allow the normal modes to be populated by an arbitrary number of particles, it is expected that the level-repulsion will disappear from the resulting system. This is intuitive, since two non-interacting identical systems will not retain the information about level-correlations that is present in one of them. This can further be explicitly checked by taking {\it e.g.}~multiple random matrix spectra and observing that the resulting full spectrum displays a Poisson level spacing distribution, as opposed to a Wigner-Dyson one. Therefore, if we consider a multi-particle SFF for our system, without introducing any further interactions, it is expected that the DRP-structure of the multi-particle SFF will disappear.\footnote{It is, in principle, possible to verify this claim by an explicit computation. However, numerically, the evaluation of the multi-particle SFF appears rather unwieldy and lacking in precision.} We hope, however, that the ``seed of chaos" which we observe in the single-particle sector, will nevertheless show up once an adequate interaction is turned on.

The above possibility brings about several intriguing questions in the general context of chaos in quantum field theories. We can formulate a QFT on a lattice: This essentially entails defining spin-like degrees of freedom at each lattice site. Each such spin degree of freedom can obey a non-trivial spectrum. On one extreme, the spectrum can be a harmonic oscillator like (or, any other integrable spectrum); on the other hand, it could be as chaotic as a random matrix spectrum. Intuitively, we expect that such a QFT on the lattice as a whole will always display integrable features, as long as the different spin degrees of freedom at different sites do not interact. Turning an interaction on, however, is a potentially interesting possibility. It is expected that for a certain class of interactions and in a certain regime of the interaction strength, the underlying single-particle spectrum physics governs the multi-particle dynamics, especially when it is chaotic. We are, however, not aware of any such precise classification or statement in the existing literature. We plan to visit this issue within a simplified lattice model in future.

\section{Acknowledgments}

We thank Roberto Emparan, Sumit Garg, Masanori Hanada, Shota Komatsu, Chethan Krishnan, Kyriakos Papadodimas, Nicholas Warner for useful discussions, conversations and feedbacks on this work. AK is partially supported by CEFIPRA 6304-3, DAE-BRNS 58/14/12/2021-BRNS and CRG/2021/004539 of Govt.~of India. AK further thanks the warm hospitality of the Department of Theoretical Physics, CERN where a large part of this work was carried out.

\appendix

\section{Appendix: Free Scalar in a BTZ Geometry}

In this appendix, we collect some basic details related to the canonical quantization of a free scalar field in a BTZ black hole geometry. The scalar field is treated in the probe limit, whose action is given by
\begin{eqnarray}
S= \int d^3 x \sqrt{-g} \left(- \frac{1}{2} g^{\mu\nu} \partial_\mu \Phi \partial_\nu \Phi  - \frac{m^2}{2} \Phi^2 \right) = \int d^3 x {\cal L} \ , 
\end{eqnarray}
The resulting equation of motion is given by
\begin{eqnarray}
\Box \Phi - m^2 \Phi = 0 \ . \label{cleom}
\end{eqnarray}
For convenience, we will consider the specific case of $m=0$. The corresponding canonical momenta are given by
\begin{eqnarray}
\pi = \frac{\partial {\cal L}}{\partial (\partial_t \Phi)} = (- g)^{1/2} g^{t\mu} \partial_\mu \Phi \ .
\end{eqnarray}
Subsequently, the following canonical commutation relations are imposed:
\begin{eqnarray}
&& \left[ \Phi(t, \vec{x} ), \Phi(t, \vec{x}') \right] = 0 = \left[\pi(t, \vec{x} ), \pi(t, \vec{x}')  \right] \ , \\
&& \left[ \Phi(t, \vec{x} ), \pi(t, \vec{x}')\right] = i \delta^{(2)}\left( \vec{x} - \vec{x}'\right) \ .
\end{eqnarray}
Finally, a natural inner product can be defined as:
\begin{eqnarray}
\left( \Phi_1, \Phi_2 \right) = - i \int_{\Sigma} \left( \Phi_1 \nabla_\mu \Phi_2^* - \Phi_2^* \nabla_\mu \Phi_1 \right) n^\mu \sqrt{\gamma} d^{2}x \ , \label{eqinner}
\end{eqnarray}
where $\Sigma$ is a spacelike hypersurface, $n^\mu$ is the corresponding normal and $\gamma$ is the induced metric on this hypersurface; furthermore $\Phi_{1,2}$ are solutions to the classical equations of motion in (\ref{cleom}). Here, we are considering real scalars, and hence $\Phi_2^* = \Phi_2$.

To define positive frequency modes, we need to choose a time-direction. Given a stationary geometry, a time-like Killing vector $K^\mu$ can be defined and correspondingly the positive modes are defined by
\begin{eqnarray}
{\cal L}_{K^\mu} f_\alpha = - i \omega_\alpha f_\alpha \ , 
\end{eqnarray}
where $\omega_\alpha$ and $f_\alpha$ are eigenvalues and eigenfunctions of the corresponding time-evolution operator which is defined by the Lie derivative along the Killing vector $K^\mu$.\footnote{Note that, for both rotating and non-rotating BTZ, we have a unique choice for a globally defined time-like Killing vector.} The eigenfunctions are normalized according to the inner product defined in (\ref{eqinner}), such that $\left(f_\beta, f_\alpha \right) = \delta_{\beta\alpha}$. The field $\Phi$ can now be expanded in terms of the positive modes, as follows:
\begin{eqnarray}
\Phi(t, \vec{x}) = \sum_\alpha \left(f_\alpha a_\alpha + f_\alpha^* a_\alpha^\dagger \right) \ ,
\end{eqnarray}
with 
\begin{eqnarray}
\left[ a_\beta , a_\alpha \right] = 0 = \left[ a_\beta^\dagger, a_\alpha^\dagger\right] \ , \quad \left[a_\alpha, a_\beta^\dagger \right] = \delta_{\alpha\beta} \ . 
\end{eqnarray}
The corresponding spectra can be obtained starting with the vacuum $a_\alpha \left | 0 \right \rangle = 0$. The corresponding number operator is defined as: $N = a_\alpha^\dagger a_\alpha$.

In our case, we have explicitly found out the solutions of the equations of motion in terms of hypergeometric functions, subject to the appropriate boundary conditions. Even though, in general, the hypergeometric functions are not orthonormal, we can construct an orthonormal basis using {\it e.g.}~the Gram-Schmidt procedure. For us, the explicit construction of this is not essential. Given the normal modes, we can define the single-particle partition function as follows:
\begin{eqnarray}
Z = {\rm Tr} \left( e^{-\beta H} {\cal P} \right)  \ , \quad {\cal P } = \delta_{N,1} \ .
\end{eqnarray}
The corresponding spectral form factor is obtained by appropriately analytically continuing this partition function.

\section{ Validity of Probe Limit} \label{probe_appendix}

Let's consider the Einstein-Hilbert action with matter field $\Phi$. The action functional is (up to boundary terms):
\begin{equation}\label{action}
    S[g;\Phi]=\int \left(\frac{1}{2\kappa}(R-2\Lambda)+\mathcal{L}_M\right)\sqrt{-g}  d^3x \ ,
\end{equation}
with Lagrangian density of matter field is given by
\begin{equation}
    \mathcal{L}_M=-\frac{1}{2}\left(g^{\alpha \beta}\partial_{\alpha}\Psi \partial_{\beta}\Psi  +\mu^2 \Psi^2  \right) \ .
\end{equation}
Einstein field equations for \eqref{action} are:
\begin{equation}
    G_{\alpha \beta}+\Lambda g_{\alpha\beta}=\kappa T_{\alpha\beta} \ ,
\end{equation}
where
\begin{equation*}
    T_{\alpha \beta}=\partial_{\alpha}{\Phi}\partial_{\beta}{\Phi}-\frac{1}{2}g_{\alpha \beta} g^{\mu \nu} \partial_{\mu}{\Phi}\partial_{\nu}{\Phi} \ .
\end{equation*}
Let's consider massless scalar field in static BTZ metric for simplicity whose metric is given by
\begin{equation}
  ds^2=-(r^2-r_H^2)dt^2+\frac{dr^2}{r^2-r_H^2}+r^2 d\psi^2 \ ,
\end{equation}
with $ \Box \Phi=0$ and $\Lambda=-1$. The energy density of scalar field as observed by a timelike observer with velocity $u^{\alpha}=\frac{1}{\sqrt{-g_{tt}}}(1,0,0)$ is:
\begin{align}
    \rho &=T_{\alpha \beta} u^{\alpha}u^{\beta} \nonumber \\
    &=T_{tt}(u^t)^2=-T_{tt}g^{tt}.
\end{align}
With $\Phi \sim e^{-i \omega t} e^{i m \psi} y(r)$, energy density is given by
\begin{align}\label{emtensor}
    \rho=& \frac{e^{2 i (m \phi -t \omega )}}{8 r^3 \left(r^2-1\right)^2}
     ( y(r)^2 \left(-4 m^2 \left(r^2-1\right) r^4+4 r^6 \omega ^2-8 r^4 \omega ^2-4 r^2 \omega ^2+1\right) \\
     &+4 r^2 y'(r)^2-4 r y(r) y'(r)))) \ . 
\end{align}
Near boundary, field goes as $y_{\text{bdry}}(r) \sim r^{-1/2}$ and density is given by
\begin{equation}
    \rho_{\text{bdry}}\sim  \left( \frac{\omega^2-m^2}{2}\frac{1}{r^4}-\frac{m^2}{2}\frac{1}{r^6}+ \ldots   \right) \ .
\end{equation}
Near horizon, field takes the following form \cite{Das:2022evy}:
\begin{equation}\label{backbdry}
    y_{\text{hor}}\sim \sqrt{r}\left( P (1-r^2)^{\frac{-i \omega}{2}}+Q e^{\pi \omega} (1-r^2)^{\frac{i \omega}{2}}   \right) \ ,
\end{equation}
where $P,Q$ are $\omega$ and $m$ dependent functions. Substituting this into \eqref{emtensor}, near horizon behaviour of energy density is given by,
\begin{equation}\label{energy1}
    \rho_{\text{hor}}\sim \frac{\omega^2}{(r^2-1)^4}\left( P Q-\frac{P^2}{2}(1-r^2)^{-i \omega}-\frac{Q^2}{2}(1-r^2)^{i \omega}   \right)+\ldots
\end{equation}
When $r \rightarrow 1$, last two terms in the above equation \eqref{energy1} are highly oscillatory. That's why to see the $m$-dependence of $\rho_{\text{hor}}$ we will only focus on the first term. Let's define,
\begin{equation}
    {\widetilde{\rho}_{\text{hor}}} =\omega^2 P \, Q \label{back1}
\end{equation}
In general $\omega$ and $m$ are not independent variable but for any small neighbourhood of $m$ we can approximate any function by a straight line with slope determined by the tangent at that point. So let's assume $\omega= s\, m$, where $s$ is the slope of $\omega=f(m)$ curve at that $m$ value. Then when $s$ is very small i.e. for larger values of $m$ (for example see Figure \ref{J_0_spectrum}) we can approximate,
\begin{equation} \label{back2}
    {\widetilde{\rho}_{\text{hor}}}  \sim 2^{-2 i m}\frac{\Gamma \left( \frac{1-im}{2} \right) }{\pi \Gamma \left(\frac{-im}{2}\right)} (1+2 e^{-m \pi}+e^{-2m\pi}) + {\cal O}(m) \sim 2^{-2 i m}\frac{\Gamma \left( \frac{1-im}{2} \right) }{\pi \Gamma \left(\frac{-im}{2}\right)}
\end{equation}
whose absolute value is $\frac{m}{2\pi} \tanh{\frac{\pi m}{2}}$, which is an increasing function of $m$. This implies for larger values of $m$, back reaction grows so we have to cut-off the spectrum at some finite $m=m_{\text{cut}}$ if we want to be in the probe limit.

\section{WKB Approximation}

Here, we will review and collect the basics of the WKB approximation, which we have used to determine an approximate spectrum. Consider the Schr\"{o}dinger equation:
\begin{eqnarray}
\frac{d^2\Psi (x)}{dx^2} - \frac{V(x)}{\hbar^2} \Psi(x) = 0 \ , \quad k(x)^2 = - V(x)^2 \ , \label{qmwkb}
\end{eqnarray}
where we have kept the $\hbar$-dependence explicit. The potential $V(x)$ is general, but we will assume that it is slowly varying, which will be made more precise momentarily. Towards finding a solution of the equation (\ref{qmwkb}), let us use an ansatz\footnote{Since the potential is slowly varying, we will assume that a plane-wave like solution exists, where the phase will now be a slowly varying function of the coordinate $x$.}:
\begin{eqnarray}
\Psi(x) = e^{\frac{i}{\hbar} \sigma(x)} \ , 
\end{eqnarray}
and the resulting equation becomes:
\begin{eqnarray}
i \hbar \sigma''(x) - \sigma'(x)^2 + k(x)^2 = 0 \ . \label{wkbansatz}
\end{eqnarray}
In the semi-classical regime, it is sensible to assume a perturbative expansion of $\sigma(x)$ in powers of $\hbar$:
\begin{eqnarray}
\sigma(x) = \sigma_0 (x) + \frac{\hbar}{i} \sigma_1(x) + \left( \frac{\hbar}{i}\right)^2 \sigma_2(x) + \ldots \ .
\end{eqnarray}
Plugging this expansion back in the equation, we obtain:
\begin{eqnarray}
&& {\cal O}(\hbar^0) : \quad \left( \sigma_0'(x) \right)^2 = k(x)^2 \quad \implies \quad \sigma_0(x) = \pm \int k(x) dx \ . \\
&& {\cal O}(\hbar):  \sigma_0''(x) + 2 \sigma_0'(x) \sigma_1'(x) = 0  \implies  \sigma_1'=\frac{\sigma_0''}{2\sigma_0'} = - \frac{k'}{2k} \implies \sigma_1 = - \frac{1}{2} \log k \ .
\end{eqnarray}
Upto this order, the wavefunction is given by
\begin{eqnarray}
\Psi(x) = \frac{c_1}{\sqrt{|k(x)|}} e^{\frac{i}{\hbar} \int k(x) dx} + \frac{c_2}{\sqrt{|k(x)|}} e^{-\frac{i}{\hbar} \int k(x) dx} \ , \label{clallow}
\end{eqnarray}
where $c_{1,2}$ are two undetermined constants  and we have chosen $k(x)^2>0$. Clearly, this corresponds to the classically accessible region for the given potential. Similarly, for the classically disallowed region, {\it i.e.}~$k(x)^2<0$, one obtains:
\begin{eqnarray}
\Psi(x) = \frac{d_1}{\sqrt{|k(x)|}} e^{\frac{1}{\hbar} \int |k(x)| dx} + \frac{d_2}{\sqrt{|k(x)|}} e^{-\frac{1}{\hbar} \int |k(x)| dx} \ , \label{clnallow}
\end{eqnarray}
where $d_{1,2}$ are the two undetermined constants.

Note that, for the perturbative expansion to work, we must require:
\begin{eqnarray}
\left| \hbar \frac{\sigma''(x)}{\left( \sigma'(x)\right)^2} \right| \equiv \left| \frac{d(\hbar/\sigma(x) )}{dx} \right| \ll 1 \ .
\end{eqnarray}
Using the leading order solution $\sigma'(x) = k(x)$ and defining a corresponding de Broglie wavelength $\lambda = 2\pi\hbar/k(x)$, the above inequality corresponds to: $|d\lambda/dx| \ll 2\pi$. Physically, this essentially implies that the de Broglie wavelength is slowly varying at the scale of one wavelength. This condition will be manifestly violated when $k(x) \to 0$, {\it i.e.}~near the classical turning points. Also, note that, the probability of finding a particle within a length scale where $V(x)$ remains approximately constant, is determined by the time-scale for which the particle spends time within such a region. Therefore, a factor of inverse momentum is expected: $|\Psi|^2 \sim k^{-1}$.

Let us now consider that the potential consists of a classically allowed region within $b\le x\le a$ and the classically forbidden regions are $x>a$ and $x<b$. In this case, the complete solution will take the form:
\begin{eqnarray}
&& \Psi(x) =  \frac{c_1}{\sqrt{|k(x)|}} e^{\frac{i}{\hbar} \int k(x) dx} + \frac{c_2}{\sqrt{|k(x)|}} e^{-\frac{i}{\hbar} \int k(x) dx} \ , \quad b \le x\le a \ ,\\
&& \Psi(x)  = \frac{d_1}{\sqrt{|k(x)|}} e^{\frac{1}{\hbar} \int |k(x)| dx}  \ , \quad x<b \ , \\
&& \Psi(x)  = \frac{d_2}{\sqrt{|k(x)|}} e^{- \frac{1}{\hbar} \int |k(x)| dx}  \ , \quad x>a \ .
\end{eqnarray}
Near the turning point, {\it i.e.}~when $V(x) \to 0$ as $x\to x_*$, let us assume that the potential is still sufficiently slowly varying that we can approximate:
\begin{eqnarray}
 V(x) =  V(x_*) + V'(x_*) (x-x_*) + \ldots = V'(x_*) (x-x_*) \ , 
\end{eqnarray}
where $x_*= a/b$, as we have chosen above. The general solution of the Schrodinger equation is given in terms of Airy functions:
\begin{eqnarray}
&& \Psi(x) = e_1 \text{Ai} \left( \left( \frac{V'(x_*)}{\hbar^2} \right)^{1/3} (x- x_* ) \right) + e_2 \text{Bi} \left( \left(\frac{V'(x_*)}{\hbar^2}\right)^{1/3} (x- x_* )  \right)  \ ,  x \sim x_* \ , \\
&& \text{Ai}(y) = \frac{1}{\pi} \int_0^\infty dt \cos\left( y t + \frac{t^3}{3} \right) \ , \\
&& \text{Bi}(y) = \frac{1}{\pi} \int_0^\infty dt \left[ \sin\left( y t + \frac{t^3}{3}\right) + {\rm exp} \left( y t - \frac{t^3}{3}\right) \right] \ , 
\end{eqnarray}
where $\text{Ai}$ and $\text{Bi}$ are the usual Airy functions and $e_{1,2}$ are two undetermined constants. Now, to fix the solution completely, we need to match the Airy functions with the corresponding oscillatory and decaying parts in the corresponding regions.

\section{Klein-Gordon to Schr\"{o}dinger} \label{KGSch}

Let us begin with the Klein-Gordon equation in a non-rotating BTZ background. For convenience, we will begin with the equation written in the coordinate of \cite{Das:2022evy}. The metric is:
\begin{eqnarray}
ds^2 = - (r^2 - r_h^2) dt^2 + \frac{dr^2}{(r^2 - r_h^2)} + r^2 d\psi^2 \ , \quad -\infty < t<\infty \ , 0<r<\infty \ , 0\le \psi < 2\pi \ .
\end{eqnarray}
The massless Klein-Gordon equation $\Box \Phi=0$ , in this background, takes the form:
\begin{eqnarray}
&& \left( r^2 -1\right)^2 \frac{d^2 \phi}{dr^2} + 2r\left( r^2 - 1\right) \frac{d\phi}{dr} + \left( \omega^2 - U(r) \right) \phi = 0 \ , \label{kgbtz} \\
&& U(r) = \left( r^2 -1\right) \left[ \frac{1}{r^2} \left( m^2 + \frac{1}{4}\right) + 1 - \frac{1}{4} \right] \ ,
\end{eqnarray}
where we have used an ansatz for the scalar field: $\Phi = \sum_{m,\omega} e^{im \psi} e^{-i\omega t} \phi(r)/\sqrt{r}$, and set $r_h=1$. Therefore every dimensionful quantity is measured in units of temperature, up to an order one constant. We will not keep track of the constant, since it is not relevant for the physics results that we are exploring.

To convert (\ref{kgbtz}), note that any second order differential equation of the following form:
\begin{eqnarray}
&& a_1(r) \phi'' + a_2(r) \phi' + a_3(r) \phi = 0 \ , \\
&& {\rm with} \quad \phi(r) = \Psi(r) g(r) \ , \quad 2 a_1(r) g'(r) + a_2(r) g(r) = 0 \ , 
\end{eqnarray}
can be recast as a Schr\"{o}dinger equation of the following form:
\begin{eqnarray}
&& \frac{d^2\Psi}{dr^2} - V(r) \Psi(r) = 0 \ , \\
&& {\rm where} \quad V(r) = - \frac{a_1(r) g''(r) + a_2(r) g'(r) + a_3(r) g(r)}{a_1(r) g(r)} \nonumber\\
&& = \frac{1}{4 a_1^2} \left[ a_2^2 - 2 a_2 a_1' + 2 a_1 \left( a_2'- 2 a_3 \right)  \right]\ . \label{schpot}
\end{eqnarray}

Similarly, an equation of the following form:
\begin{eqnarray}
A(z) F''(z) + B(z) F'(z) + C(z) F(z) = 0 \ , 
\end{eqnarray}
can be recast into a Schr\"{o}dinger equation of the following form:
\begin{eqnarray}
&& \frac{d^2\Psi(z)}{dz^2} - V(z) \Psi(z) = 0 \ , \quad F(z) = \Psi(z) g(z) \ , 2 A g' + B g =0 \ ,  \\
&& {\rm where} \quad V(r) = \frac{1}{4 A^2} \left[ B^2 - 2 B A' + 2 A \left( B' - 2 C \right)  \right ] \ . \label{schpotJ}
\end{eqnarray}
%


\end{document}